\documentclass[%
floatfix,
 reprint,
 amsmath,amssymb,
 aps,
]{revtex4-2}

\usepackage{graphicx}
\usepackage{dcolumn}
\usepackage{bm}

\usepackage{graphicx}
\usepackage{color}
\usepackage{ulem}
\usepackage{dsfont}
\usepackage{bm}
\usepackage[utf8]{inputenc}
\usepackage[english]{babel}
\usepackage{feynmp}
\usepackage{bbold}
\usepackage{tensor}
\usepackage{tikz}

\begin{document}

\preprint{APS/123-QED}

\title{Klein paradox between transmitted and reflected Dirac waves on Bour surfaces}

\author{V\'ictor A.  \surname{Gonz\'alez-Dom\'inguez}}
\affiliation{Instituto de Investigaci\'on e Innovaci\'on en Energ\'ias Renovables, Universidad de Ciencias y Artes de Chiapas, Libramiento Norte Poniente No. 1150, C.P. 29039, Tuxtla Guti\'errez, Chiapas, M\'exico}

\author{Juan A.  \surname{ Reyes-Nava}}
\affiliation{Instituto de Investigaci\'on e Innovaci\'on en Energ\'ias Renovables, Universidad de Ciencias y Artes de Chiapas, Libramiento Norte Poniente No. 1150, C.P. 29039, Tuxtla Guti\'errez, Chiapas, M\'exico}

\author{Pavel \surname{Castro-Villarreal}}
\email[]{pcastrov@unach.mx}
\thanks{author to whom correspondence should be addressed.}
\affiliation{Facultad de Ciencias en F\'isica y Matem\'aticas,
Universidad Aut\'onoma de Chiapas, Carretera Emiliano Zapata, Km. 8, Rancho San Francisco, 29050 Tuxtla Guti\'errez, Chiapas, M\'exico}

\begin{abstract}
It is supposed the existence of a curved graphene sheet with the geometry of a Bour surface $B_{n}$, such as the catenoid (or helicoid), $B_{0}$, and the classical Enneper surface, $B_{2}$, among others. In particular, in this work, the propagation of the electronic degrees of freedom on these surfaces is studied based on the Dirac equation. As a consequence of the polar geometry of $B_{n}$, it is found that the geometry of the surface causes the Dirac fermions to move as if they would be subjected to an external potential coupled to a spin-orbit term. The geometry-induced potential is interpreted as a barrier potential, which is asymptotically zero. Furthermore, the behaviour of asymptotic Dirac states and scattering states are studied through the Lippmann-Schwinger formalism. It is found that for surfaces $B_{0}$ and $B_{1}$, the total transmission phenomenon is found for sufficiently large values of energy, while for surfaces $B_{n}$, with $n\geq 2$, it is shown that there is an energy point $E_{K}$ where Klein's paradox is realized, while for energy values $E\gg E_{K}$ it is found that the conductance of the hypothetical material is completely suppressed, $\mathcal{G}(E)\to 0$.
\end{abstract}

\maketitle


\section{Introduction}

One of the intriguing properties of graphene, among many others \cite{Novoselov},  is that the charge carriers can be described by quasiparticles with the same behaviour as relativistic Dirac fermions at the low-energy regime \cite{Novoselov2}. This fact had been theoretically predicted when it was shown that Dirac's field theory emerges \cite{Gordon,Mele} from Wallace's tight-binding model \cite{Wallace}. These characteristics of graphene allowed establishing an analogy with relativistic quantum phenomena \cite{Geim2007}. Even more so, it is possible to think of graphene as the mother of other graphitic materials, since it can fold up to form fullerenes \cite{Kroto,Lamb}, roll up into carbon nanotubes \cite{Lijima} and stack up to shape graphite. Furthermore, using concepts from geometry and topology new carbon nanostructured curved materials can be created with new properties \cite{Terrones3, Melinon2021}. Indeed, curved carbon materials were proposed more than a decade before the advent of groundbreaking graphene \cite{Terrones}. Although there is still no experimental synthesis of these nanostructured materials, there is a good expectation \cite{Terrones2} that this will be so with the technological advancement \cite{Braun2018, Tanabe2020, Melinon2021}. 

The possibility of studying quantum phenomena on curved space-times was also raised \cite{Vozmediano} with the logical implication to explore gravitational analogues phenomena in tabletop experiments \cite{Iorio2013, 10.3389/fmats.2014.00036}, for instance,  through the conformal-gauge symmetry encoded in the $2+1$ Dirac theory \cite{Iorio, Cvetic}, the intrinsic curvature of the graphene sheet gives up a general relativistic description of fermionic degrees of freedom, whereas the electronic properties of a sheet with a shape of a Beltrami trumpet \cite{Gallerati-2021} is interpreted as the Unruh-Hawking effect \cite{Iorio-Limbiagie, Iorio2014, Tommaso-2020}, forasmuch under a specific external magnetic field the space-time metric is described by a Zermelo optical metric which is conformally equivalent to the BTZ black hole metric \cite{Cvetic}, where also Hawking-radiation phenomena can be explored \cite{Kandemir}.  Likewise, using a variation of the hopping parameters in the tight-binding model an emergent Horava gravity arises \cite{Volovik}. Additionally, a simulation of quantum gravity analogues can be achieved when chiral symmetry is broken; that is, when it is taking into account trigonal warping of the electronic spectrum \cite{Alonso-Astorga, Iorio_2019}. From the condensed matter perspective, the Dirac equation in the curved space is a natural model for studying the electronic properties of the graphene sheet when undulations and topological defects are present, for instance, one can address the problem of impurities and topological defects  \cite{Gallerati-2022,Chaves-2014,Chaves-2017}; in fact, the QFT formulation on the curved space might also describe the external strain acting on the material \cite{DeJuan2012}. Furthermore, the corrugations on a curved sheet of graphene give up the appearance of a pseudo-magnetic field \cite{Levy2010, Vozmediano2010}, which has the remarkable property to be proportional to the Ricci scalar curvature \cite{Arias, Villarreal}. 

Geometric and topology effects on the behaviour of quantum states have been of quite interest in the community for a while. For example, by formulating Shr$\ddot{\rm o}$dinger quantum mechanics on curves and surfaces using {\it confining potential formalism} \cite{DaCosta1981, Jaffe, Ferrari2008}  the quantum states are analyzed for particles confined to a helix, catenary, helicoid (or catenoid) \cite{Lima}, also a piece of evidence was found for reminiscences of an analogous quantum Hall effect when bending a strip helically \cite{Dandaloff-2009,Dandoloff-2004}. At the same time motivated by the physics behind a wormhole spacetime in general relativity \cite{Thorne}, the catenoid was shown to be the analogous wormhole model in 2+1, while the external electric and magnetic background fields on the catenoid material give up bound states around the bridge, and produce modified Landau levels \cite{DaCosta}. From the intrinsic perspective, \cite{DeWitt},  a geometry-induced potential for Dirac fermions is deduced from the intrinsic geometry of the helicoid, resulting in the emergence of a pseudo-electric field near the potential minima giving rise to a chiral separation on the opposite rims of the helicoid \cite{Atanasov-Saxena}. 

The helicoid and catenoid are isometric surfaces that belong to the family of minimal surfaces, that are, those surfaces that minimize area or equivalently those such that mean curvature is zero at each point on the surface, implying a negative Ricci scalar curvature. Also, minimal surfaces are solutions of the Willmore shape equation used to describe the conformation of soft surfaces in biophysics \cite{PhysRevE.76.011922}.
These surfaces were proposed three decades ago as representations of new nanostructured allotropes of carbon obtained by decorating a minimal surface with carbon arrangement such as the triply periodic minimal surface by considering the inclusion of octagon rings in the carbon lattice structure \cite{Terrones}. Here, we assume that minimal surfaces are stable nanostructured graphitic shapes \cite{Terrones3}; particularly, we focus on a subset of this family known as Bour surfaces \cite{Whittemore} where the catenoid, helicoid or the classical Enneper surface belong, among other surfaces, all characterized by their polar symmetry. As a consequence of the symmetry and  the  Weierstrass-Enneper representation \cite{Ulrich} we can simplify the Dirac equation to study the electronic degrees of freedom on these surfaces. Furthermore, we determine a geometry-induced potential for the Bour surface that generalizes that found for the helicoid \cite{Atanasov-Saxena}. In particular, we study the propagation of Dirac waves using the Lippmann-Schwinger equation along the latitudinal lines of the hypothesised material to address the problem of elucidating the role of geometry in the scattering states, and the conductance of the material.

This article is organized as follows. In section (\ref{sectII}), the non-coordinates basis notation and the Weierstrass-Enneper representation for a minimal surface are introduced. These elements are necessary so that in the section (\ref{sectIII}) we write down the Dirac equation in spacetime $\mathbb{M}=\mathbb{R}\times \Sigma$, where $\Sigma$ is a surface of Bour. In section (\ref{sectIV}), the Dirac equation is rewritten so that it can be interpreted as the equation for a Dirac fermion subjected to a repulsive potential coupled to a spin-orbit term. Furthermore, the asymptotic Dirac states are determined, which are used as initial states in the section (\ref{sectV}) to determine the scattered states based on the Lippmann-Schwinger formalism. Furthermore, in section (\ref{sectVI}) the transmittance and reflectance coefficients are calculated using the N$\ddot{\rm o}$ether current. In the section (\ref{sectVII}), conclusions and perspectives are presented, also in \ref{AppA}-\ref{AppC} appendices have been added.

\section{Geometrical preliminaries}\label{sectII}

This section introduces geometrical preliminaries suitable to set up the analysis of electronic degrees of freedom for curved Dirac materials performed here. Particularly, it introduces the tetrad formalism for the geometry of a $2+1$ space-time, which allows us to write down the Dirac equation on a curved space-time associated with the curved Dirac material. Additionally, it is presented the Weierstrass-Enneper representation of a minimal surface, $\Sigma$, embedded in $\mathbb{R}^{3}$; in particular, it is introduced the Bour sub-family of minimal surfaces.  These surfaces are introduced since we shall analyze the electronic propagation on materials associated with the space-time $\mathbb{R}\times \Sigma$.

\subsection{Tetrad formalism for a $2+1$ space-time}

Let us start introducing first local coordinate bases for a $2+1$ space-time geometry $\mathbb{M}$. For the tangent space $T_{p}\mathbb{M}$, let the set $\{\partial_{\mu}\}$ be a local coordinate basis, whereas the set $\{dx^{\mu}\}$ the corresponding  basis for the cotangent space $T^{*}_{p}\mathbb{M}$. Here,   $p\in \mathbb{M}$ and the bi-orthogonality condition $dx^{\mu}(\partial_{\nu})=\delta^{\mu}_{\nu}$ is satisfied. The greek indices $\mu$ split in the chosen local coordinate patch. 

In the following, it is presented a non-coordinate basis for the tangent space as the set $\{\hat{e}_{A}=e^{\mu}_{A}\partial_{\mu}\}$, where the capital latin indices $A$ are global indices $A=0,1,2$, and  the coefficients sort up in a matrix structure $E$ build up an element of $GL(3, \mathbb{R})$. These coefficients are also called  vielbeins that are attached to a local patch of $\mathbb{M}$.  The non-coordinate basis are defined in such a way that it diagonalizes the metric tensor $g=g_{\mu\nu}dx^{\mu}\otimes dx^{\nu}$, that is $g(\hat{e}_{A}, \hat{e}_{B})=\eta_{AB}$, where  $\eta_{AB}={\rm diag}\left(-1,1,1\right)$ is the Minkowski space-time metric. This means that one can written down the metric tensor components as $g_{\mu\nu}=\eta_{AB}e^{A}_{\mu}e^{B}_{\nu}$, where $e^{A}_{\mu}$ are the elements of the inverse matrix $E^{-1}$. Additionally, the corresponding non-coordinate basis for the cotangent space is defined as the set $\{\hat{\theta}^{A}=e^{A}_{\mu}dx^{\mu}\}$. Now, the tensor metric can be written as $g=\eta_{AB}\hat{\theta}^{A}\otimes \hat{\theta}^{B}$, that is clearly diagonal.

Now, let us introduce a connection one-form $\omega\indices{^{A}_{B}}=\hat{\Gamma}^{A}_{CB}\hat{\theta}^{C}$, where  $\hat{\Gamma}\indices{^{A}_{BC}}$ are the coefficients of the affine connection, $\nabla$, defined through the equation $\nabla_{A}\hat{e}_{B}=\hat{\Gamma}\indices{^{C}_{AB}}\hat{e}_{C}$. The connection one-form  encodes geometrical information of $\mathbb{M}$ through the Maurer-Cartan structure equations which are given by 
\begin{eqnarray}
d\hat{\theta}^{A}+\omega\indices{^{A}_{B}}\wedge \hat{\theta}^{B}&=&\mathcal{T}^{A},\label{Maurer1}\\
d\omega\indices{^{A}_{B}}+\omega\indices{^{A}_{C}}\wedge\omega\indices{^{C}_{B}}&=&\mathcal{R}\indices{^{A}_{B}}, 
\end{eqnarray}
where $\mathcal{T}^{A}$ and $\mathcal{R}\indices{^{A}_{B}}$ are the torsion and Riemann curvature of the manifold, respectively. Further, the connection one-form written in the local coordinate basis looks like $\omega\indices{^{AB}}=\omega\indices{_{\alpha}^{AB}}dx^{\alpha}$, where $\omega^{AB}=\delta^{BC}\omega\indices{^{A}_{C}}$. Furthermore, if one ask for a Levi-Civita  affine connection $\nabla$, one has  the metric compatibility condition $\nabla_{X}g=0$ for any vector field $X$. This implies the conditions $\omega_{AB}=-\omega_{BA}$ and $\mathcal{T}^{A}=0$ \cite{Nakahara:2003nw}.  

\subsection{Weierstrass-Enneper representation for a  minimal surface}

Minimal surfaces are mathematical surfaces embedded in the space corresponding to minimizing area-surface, as well as minimizing of the Willmore energy \cite{Ulrich}.  The geometry of these surfaces is determined using the Weierstrass-Enneper (WE) representation  following \cite{Ulrich}. The WE representation of a minimal surface can be cast in terms of the mapping  ${\bf X}:\Omega\subset\mathbb{C}\to \Sigma\subset\mathbb{R}^{3}$ defined by the embedding functions 
\begin{eqnarray}
{\bf X}\left(\omega\right)={\bf X}_{0}+\Re\int_{\omega_{0}}^{\omega}\boldsymbol{\Phi}\left(\underline{\omega}\right)d\underline{\omega},\label{WErep}
\end{eqnarray}
with $\omega\in \Omega$, where $\Omega$  is a simply connected domain $\Omega$. The differential volume $d\underline{\omega}$ represents an appropriate measure for $\Omega$; $\Re$ and $\Im$ denotes the real and imaginary part, respectively. The function $\boldsymbol{\Phi}\left(\omega\right)$ can be written in terms of a holomorphic function $\mathcal{F}\left(\omega\right)$ as
\begin{eqnarray}
\boldsymbol{\Phi}\left(\omega\right)=\left(\left(1-\omega^2\right)\mathcal{F}\left(\omega\right), i\left(1+\omega^2\right)\mathcal{F}\left(\omega\right), 2\omega \mathcal{F}\left(\omega\right)\right),
\end{eqnarray}
where $\mathcal{F}\left(\omega\right)$ is called Weierstrass function. The Gauss map using this WE representation is given by the normal vector field ${\bf N}\left(\omega\right)=\left(2\Re \omega, 2\Im \omega, \left|\omega\right|^2-1 \right)/(1+\left|\omega\right|^2)$. Additionally, by
taking the real and imaginary part of $\omega=u+i v\in \mathbb{C}$, respectively, it can be defined a  local patch with local coordinates $\{u, v\}$. 

Given an specific Weierstrass function $\mathcal{F}(\omega)$ one can determine the embedding functions ${\bf X}(\omega)$ which allows us to determine the whole exitrinsic and intrinsic geometry of $\Sigma$. For instance, the main feature of a minimal surface is the vanishing mean curvature,  $H=0$.  Furthermore, the intrinsic geometry of a minimal surface is  described in terms of the metric tensor,  introduced, here,  through the square of the line element $ds^2=g_{ab} d\xi^{a} d\xi^{b}$, where the metric tensor components, $g_{ab}$, are calculated  using the equation $g_{ab}=\partial_{a}{\bf X}\cdot \partial_{b}{\bf X}$, with indices $a, b=u, v$. Using the above embedding functions of WE representation, it can be shown that
$ds^2=\Lambda^2(\omega)\left|d\omega\right|^2$,
where $\left|d\omega\right|^2=du^2+dv^2$, and the conformal factor, $\Lambda^2(\omega)$, is given by $\Lambda^2(\omega)=\left|\mathcal{F}\left(\omega\right)\right|^2\left(1+\left|\omega\right|^2\right)^2$. Thus, WE representation gives already local isothermal coordinates that always exist in a two-dimensional manifolds \cite{Chern1955AnEP}. In addition, the Gaussian curvature is given by 
$K=-1/\left|\mathcal{F}\left(\omega\right)\right|^2\left(1+\left|\omega\right|^2\right)^4$
for points $\omega\in\Omega^{\prime}$, where  the set of regular points is given by $\Omega^{\prime}=\{\omega\in\Omega: \mathcal{F}\left(\omega\right)\neq 0\}$.

\begin{figure}[ht]
    \centering
    \includegraphics[scale=0.3]{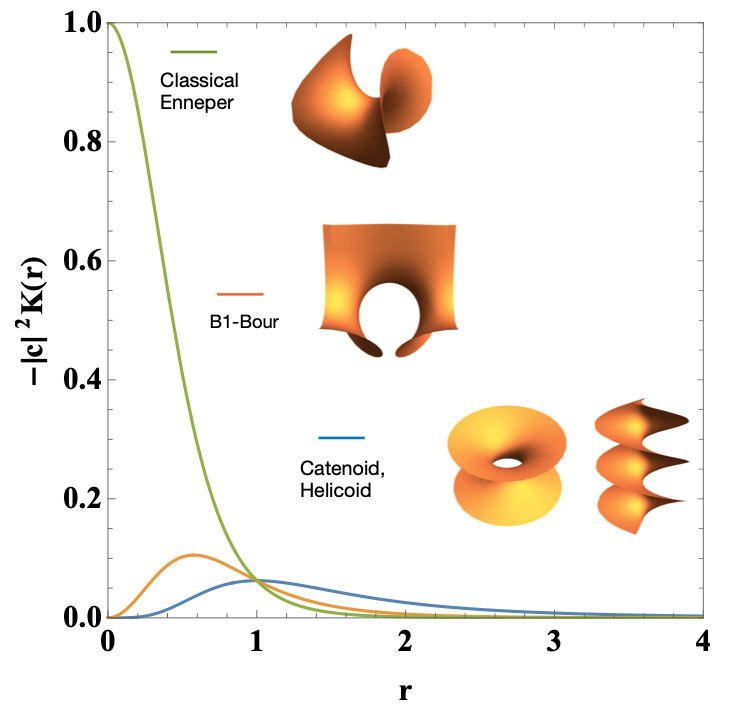}
    \caption{Negative of Gaussian curvature (Eq. (\ref{BourCurvature})) vs radial coordinate $r$ for the  cases $n=0,1,2$. The figure also shows examples of minimal Bour's surfaces from the top to the bottom: classical Enneper, $B_{1}$-Bour, Catenoid  and Helicoid surfaces, respectively. All the surfaces inset have finite value of curvature. }
    \label{fig1}
\end{figure}
In the following, we  focus in a sub-family of minimal surfaces known as Bour's minimal surfaces \cite{Ulrich, Whittemore} defined through the Weierstrass function $\mathcal{F}\left(\omega\right)=c\omega^{n-2}$, where $c\in\mathbb{C}$ and $n\in\mathbb{R}$. In this sub-family belong the catenoid with $n=0$ and $c=R_{0}/2$, being $R_{0}$ is the radius of the neck; the helicoid with $n=0$ and $c=i\alpha$ being $\alpha$ the pitch of the helicoid, and  the Enneper surface with $n=2$, $c=1$. Also, it is known that $n$ and $-n$ represents the same Bour surface, thus it is enough to consider the cases $n\geq 0$ \cite{Whittemore}.  Notice that in the case of the Bour's surfaces the conformal factor depends just on the norm $\left|\omega\right|=\sqrt{u^2+v^{2}}$, thus it is  convenient to use polar  coordinates, $(r, \theta)$, defined as usual $r=\left|\omega\right|$ and $\theta=\arctan\left(v/u\right)$. The conformal factor is just given by  
\begin{eqnarray}
\Lambda(r)=\left|c\right| r^{n-2}\left(1+r^{2}\right), \label{ConformalFactorBour}
\end{eqnarray}
where we recall $c\in\mathbb{C}$ and $n\geq 0$ are parameters that gives a specific Bour's surface. Now the Gaussian curvature of these surface is given by 
\begin{eqnarray}
K(r)=-\frac{1}{\left|c\right|^2 r^{2(n-2)}\left(1+r^2\right)^{4}}\label{BourCurvature}.
\end{eqnarray}
By inspection, one can see that the Gaussian curvature is finite for $r\in \mathbb{R}$ for the cases $n=0, 1,2$ (see figure (\ref{fig1})), while for $n>2$ all the $B_{n}$-Bour surfaces have a singular curvature at $r=0$. In addition, since Gaussian curvature has units of the inverse square of length thus $\left|c\right|$ gives a natural length unit for the Bour surfaces. Note that the catenoid and helicoid have the same Gaussian curvature since these surfaces are isometric with each other.

\section{Electronic states: Dirac field on the curved space-time $\mathbb{M}=\mathbb{R}\times\Sigma$}\label{sectIII}
In this section, we introduce the Dirac model on a space-time geometry with the global structure $\mathbb{M}=\mathbb{R}\times\Sigma$, where $\Sigma$ is a minimal surface. Later on, $\Sigma$ is specified as a specific member of the Bour minimal surface family. 

\subsection{Dirac equation on the curved space-time $\mathbb{M}=\mathbb{R}\times\Sigma$}\label{DiracEq}
 The starting point is the Dirac equation $i\underline{\gamma}^{\alpha}\boldsymbol{\nabla}_{\alpha}\Psi=0$, defined on a $2+1$ space-time $\mathbb{M}$, where $\underline{\gamma}^{\alpha}\left(x\right)=\gamma^{A}e^{\alpha}_{A}\left(x\right)$ for $x\in\mathbb{M}$, where $e^{\alpha}_{A}$ are the vielbeins introduces in the previous section, and being  $\gamma^{A}$ the Dirac matrices that satisfy the Clifford algebra 
\begin{eqnarray}
\left\{\gamma^{A}, \gamma^{B}\right\}=2\eta^{AB}\mathbb{1},
\label{CliffordAlgebra}
\end{eqnarray}
where $\gamma^{A}$ have range $2$ and $\mathbb{1}$ is the unit diagonal matrix. A suitable representation of the Dirac matrices in $2+1$ space-time dimension that satisfy the Clifford algebra (\ref{CliffordAlgebra}) is given by the matrices $\gamma^{0}=-i\sigma_{3}$, $\gamma^{1}\gamma^{0}=\sigma_{1}$ and $\gamma^{2}\gamma^{0}=\sigma_{2}$, where $\sigma_{i}$,  are the standard Pauli matrices, with $i=1,2,3$.  In addition, $\boldsymbol{\nabla_{\alpha}}$ is the covariant derivative for the spinor representation of the Lorentz group $SO(2,1)$ acting on the Dirac spinors as $\boldsymbol{\nabla}_{\alpha}\Psi$, where  $\boldsymbol{\nabla}_{\alpha}=\partial_{\alpha}+\Omega_{\alpha}$, being $\Omega_{\alpha}=\frac{1}{8}\omega^{AB}_{\alpha}\left[\gamma_{A}, \gamma_{B}\right]$ the spin connection, and $\omega\indices{_{\alpha}^{AB}}$ are the components of the Maurer-Cartan one-form connection defined in the previous section.

\subsection{Polar coordinates.} The metric of the space-time $\mathbb{R}\times \Sigma$ using polar coordinates  is written through the square of the line element as follows 
\begin{eqnarray}
ds^{2}=-v_{F}^{2}dt^2+\Lambda^{2}(r, \theta)\left(dr^2+r^2 d\theta^2\right).\label{SpaceTimePolar}
\end{eqnarray}
 The local indices in this case can be split as $\alpha=t, r,\theta$.  From the metric (\ref{SpaceTimePolar}) one can easily read $\hat{\theta}^{0}=v_{F}dt$, $\hat{\theta}^{1}=\Lambda dr$ and $\hat{\theta}^{2}=\Lambda r d\theta$, from where one can extract the components of the vielbeins $e^{A}_{\mu}$\footnote{Similar procedure can be done in the cartesian coordinates (see appendix (\ref{AppC}) ).}. Now, from the Maurer-Cartan equation (\ref{Maurer1}) and the torsionless condition one can obtain $d\hat{\theta}^{0}=0$, and 
\begin{eqnarray}
d\hat{\theta}^{1}+\frac{\Lambda_{\theta}}{\Lambda^{2}r}\hat{\theta}^{1}\wedge \hat{\theta}^{2}&=&0\label{MCCatt1}\\
d\hat{\theta}^{2}+\frac{\left(\Lambda r\right)_{r}}{\Lambda^{2}r}\hat{\theta}^{2}\wedge \hat{\theta}^{1}&=&0\label{MCCatt2},
\end{eqnarray}
where $\left(X\right)_{r}\equiv\partial_{r}X$. Now, from the equation (\ref{MCCatt1}) one can deduce $\omega\indices{^{1}_{0}}=0$ and $\omega\indices{^{1}_{2}}=\frac{\Lambda_{\theta}}{\Lambda^2 r}\hat{\theta}^{1}+X\hat{\theta}^{2}$ for some local function $X$, whereas from (\ref{MCCatt2}) one can deduce that  $\omega\indices{^{2}_{0}}=0$ and $\omega\indices{^{2}_{1}}=\frac{\left(\Lambda r\right)_{r}}{\Lambda^2 r}\hat{\theta}^{2}+\tilde{X}\hat{\theta}^{1}$. Now, we use the metric condition, $\omega^{AB}=-\omega^{BA}$, thus one can determine $X$ and $\tilde{X}$, turning that the only non-zero components of the connection one-form are 
\begin{eqnarray}
\omega^{12}=-\omega^{21}=\frac{\Lambda_{\theta}}{\Lambda^2 r}\hat{\theta}^{1}-\frac{\left(\Lambda r\right)_{r}}{\Lambda^2 r}\hat{\theta}^{2}.
\end{eqnarray}
These components expressed in local coordinates are given by
$\omega\indices{_{r}^{12}}=-\omega\indices{_{r}^{21}}=\frac{1}{r}\partial_{\theta}\log\Lambda$ and $\omega\indices{_{\theta}^{12}}=-\omega\indices{_{\theta}^{21}}=-\frac{1}{\Lambda}\partial_{r}\left(\Lambda r\right)$.  Consequently, the spin connection $\Omega_{\alpha}$ is given simply by $\Omega_{t}=0$, $\Omega_{r}=\frac{i}{2}\frac{1}{r}\partial_{\theta}\left(\log\Lambda\right)\sigma_{3}$, and $\Omega_{\theta}=-\frac{i}{2}\frac{1}{\Lambda}\partial_{r}\left(\Lambda r\right)\sigma_{3}$.

Now, we use all these information in order to write down an explicit expression for the Dirac equation in polar coordinates. Denoting the $2+1$ Dirac spinor by $\Psi$ and performing now the transformation $\Psi=r^{-\frac{1}{2}}\Lambda^{-\frac{1}{2}}\Phi$, we are able to show that the Dirac equation can be written as 
\begin{eqnarray}
i\hbar\partial_{t}\Phi=-i\frac{\hbar v_{F}}{\Lambda}\left(\sigma_{1}\partial_{r}\Phi+\frac{1}{r}\sigma_{2}\partial_{\theta}\Phi\right).
\label{DiracPolar}
\end{eqnarray}
The Dirac equation in these coordinates $(r,\theta)$ shall be 
particularly useful in the case when the conformal factor $\Lambda$ depends on one of the coordinates. For instance, for the surfaces considered here, that is, for the Bour's minimal surface family, where $\Lambda$ depends just on the variable $r$. 

\section{Geometry-induced potential and asymptotic states on the Bour's minimal surfaces family}\label{sectIV}

In this section, it is determined the geometry-induced potential and the Dirac asymptotic states on the Bour's minimal surface. These states are defined as the solutions of the Dirac equation for $r\to \infty$.

\subsection{Dirac fermions under effective potential}

 Our starting point is the Dirac equation on polar coordinates deduced above (\ref{DiracPolar}), where the conformal factor $\Lambda\left(r\right)$ is given just by (\ref{ConformalFactorBour}).  Since the conformal factor depends just on one of the coordinates one can make a further change of variable, using the transformation 
\begin{eqnarray}
x_{n}(r)=\int \Lambda(r)dr=\left|c\right|\left[\frac{r^{n-1}}{n-1}+\frac{r^{n+1}}{n+1}\right]\label{ChVar}
\end{eqnarray}
for $n\neq 1$, while for $n= 1$ the appropiate change of variable is   $x_{1}(r)=\left|c\right|\left(\log r+\frac{1}{2}r^{2} \right)$. One can verify that these transformation are injectives maps, thus one can guarantee the existence of their corresponding inverse functions $r=r(x)$, where  $x$ would be defined in an appropriate domain $\mathcal{D}_{x}$. Using this variable the Dirac equation (\ref{DiracPolar}) can be simplified as 
\begin{eqnarray}
i\hbar\partial_{t}\Phi=v_{F}\sigma_{1}\hat{p}_{x}\Phi+v_{F}\sigma_{2}V(x)\hat{\ell}_{\theta}\Phi,\label{DiracEqBour}
\end{eqnarray}
where $\hat{p}_{x}=-i\hbar \partial_{x}$ is a linear momentum operator and $\hat{\ell}_\theta=-i\hbar \partial_{\theta}$ is a two-dimensional angular momentum operator. Noticeably, the second term of this equation can be interpret as an effective potential; although this term is entirely coming from the intrinsic geometry of the surface, thus this potential is a geometry-induced potential. The effective potential is given by 
\begin{eqnarray}
V(x)=\frac{1}{r(x)\Lambda(r(x))}\label{potential}.
\end{eqnarray}
This potential generalizes the effective potential found in  \cite{Atanasov-Saxena} for the helicoid.

Let us note that for those cases with $n\geq 2$, such as the classical Enneper surface, the domain of the variable $x$ is $\mathbb{R}^{+}$ (see Eq. (\ref{ChVar})). Thus it is useful to define an extension to the whole reals  defining $U(x)=V(x)$ for $x\geq 0$ and $U(x)=V(-x)$ for $x\leq 0$ as an extension for $x\in \mathbb{R}$.  This construction makes the potential symmetric, and the Dirac equation for this extension  turns out to be $i\hbar\partial_{t}\Phi=v_{F}\sigma_{1}\hat{p}_{x}\Phi+v_{F}\sigma_{2}U(x)\hat{\ell}_{\theta}\Phi$, which clearly reduces to (\ref{DiracEqBour}) for $x\geq 0$.

\begin{figure}[ht]
    \centering
    \includegraphics[scale=0.37]{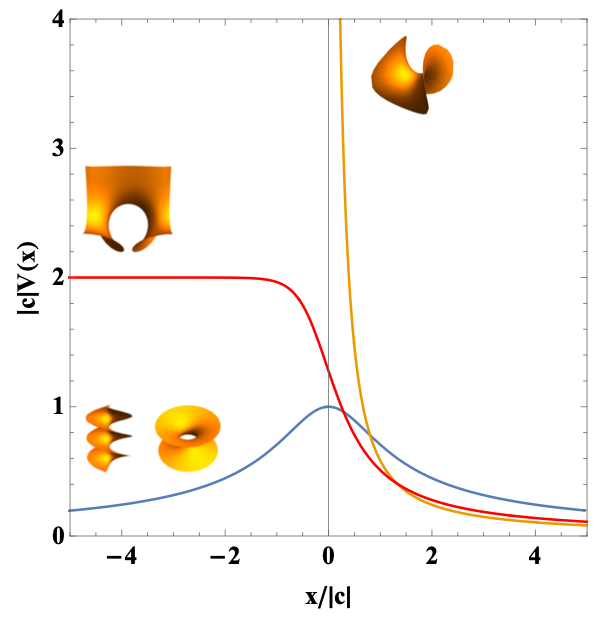}
    \caption{Effective potential (Eq. (\ref{potential})) vs coordinate $x$ (\ref{ChVar}) for the classical Enneper ($n=2$), $B_{1}$-Bour, $n=1$, and Catenoid (Helicoid)  ($n=0$).}
    \label{fig22}
\end{figure}

Now, we deduce main characteristics of the effective potential $V(x)$. Using the conformal factor (\ref{ConformalFactorBour}) and the change of variable $x(r)$ it can be shown that near $x\approx 0$ for the  catenoid (or helicoid) surface with $n=0$, the potential is $V(x)\approx c$, where $c$ is a positive constant;  for $B_{1}$-Bour surface $n=1$, the potential is linear $V(x)\simeq m x+b$, with a negative slope $m$ and a positive  $y$-intercept $b$;  and for any Bour surface with $n>1$ the effective potential behaves as $V(x)=c/((n-1)x)$, for some positive constant $c$. Therefore, the potential $V(x)$ works as a scattering potential. Furthermore, here, we are interested in the states faraway from the center $x=0$ to use them as initial states that  propagate along the surface. Additionally,  one can verify that the effective potential, $V(x)$, vanishes for $x\to\infty$ for all $n\geq 0$.  In figure (\ref{fig22}) it is drawn the geometry-induced potential for the first three cases $n=0,1,2$. 
\paragraph{Example $B_{0}$ (Catenoid/Helicoid).} Let us note that for example in the case of the catenoid (or helicoid) $n=0$ and  $\left|c\right|=R_{0}/2$, one has  $x=\frac{R_{0}}{2}\left(r-r^{-1}\right)$, where the domain is the whole $\mathbb{R}$ and recall $R_{0}$ is the radius of the neck of the catenoid. Using the expression for the conformal factor (\ref{ConformalFactorBour}) and the generalized effective  potential (\ref{potential}) it is straightforward to show that $V(x)=1/\sqrt{x^{2}+R^{2}_{0}}$, which satisfies the qualitative characteristics deduced above. Note that this effective potential is exactly the same found in \cite{Atanasov-Saxena}

\paragraph{Example ($B_{1}$-Bour surface).} Let us note that for example in the case of $B_{1}$-Bour surface $n=1$ and  choosing $c=1$, one has  $x=\log r+\frac{1}{2}r^2$, where the domain is the whole $\mathbb{R}$. One can express $r$ in terms of $x$ using the principal value of the  Lambert W function  $r^2=W\left(e^{2x}\right)$.
Using the expression for the conformal factor (\ref{ConformalFactorBour}) and the generalized effective  potential (\ref{potential}) it is straightforward to show that $V(x)=1/\left(1+W\left(e^{2x}\right)\right)$, which satisfies the qualitative characteristics deduced above.

\paragraph{Example $B_{2}$ (Enneper surface).} Now, for the Enneper classical surface $n=2$ using $c=1$,  one has  
\begin{eqnarray}
x=r+\frac{r^{3}}{3}.\label{Polynomial3rd}
\end{eqnarray} 
Let us note that contrary to the previous examples in this case the domain of this variable $x\in \mathbb{R}^{+}$. Using the expression for the conformal factor (\ref{ConformalFactorBour}) and the generalized effective  potential (\ref{potential}) is $V(x)=1/(r(x)+r^{3}(x))$, thus one need to found the positive root of a third-order polynomial (\ref{Polynomial3rd}). The effective potential in terms of $x$ is given explicitly  by
\begin{eqnarray}
V(x)=\frac{1}{3x+2^{\frac{2}{3}}\left(R^{\frac{1}{3}}_{-}(x)-R^{\frac{1}{3}}_{+}(x)\right)}\label{EnneperPot}
\end{eqnarray}
where $R_{\pm}(x)=\sqrt{9x^2+4}\pm 3x$. Note that series expansion of $R^{1/3}_{\pm}(x)$ around $x\simeq 0$ is $R^{1/3}_{\pm}(x)\simeq 2^{\frac{1}{3}}\pm 2^{-\frac{2}{3}}x$. Thus the effective potential near $x=0$ behaves similar to a Columbic type potential,  $V(x)\simeq\frac{1}{x}$, while it is clear that for large values of $x$, that is for $x\to\infty$ the potential $V(x)$ vanishes as was shown qualitatively above. 

\subsection{Asymptotic Dirac states on the Bour's minimal surfaces}

The previous analysis allows us to justify that  for the asymptotic states one can neglect the second term of (\ref{DiracEqBour}). Thus the equation reduces to a $1+1$-Dirac equation $i\partial_{t}\Phi=v_{F}\sigma_{1}\hat{p}_{x}\Phi$. For the solutions of this equation we propose the spinor solution as $\Phi(x, \theta, t)=e^{i kx -i \frac{E t}{\hbar}}f(\theta)v$, where  the function $f(\theta)$ is an arbitrary non-zero periodic function in the azimuthal angle $\theta$, and $v$ is a vector that acquires the pseudo-spin character of the spinor. By imposing the one-dimensional Dirac equation, the dispersion relation turns out to be $E=\pm \hbar v_{F}\left|k\right|$,  and $v$ satisfies the equation $\hat{h}v=\pm\frac{1}{2}v$, where $\pm$ it refers a positive and energy Dirac states, where 
 \begin{eqnarray}
 \hat{h}=\frac{1}{2}\sigma_{1}\frac{k}{\left|k\right|},
 \end{eqnarray}
 is the one-dimensional analogue of the helicity operator. For positive energy value the pseudo-spin states are given by $v_{\uparrow}$ ($v_{\downarrow}$) for positive (negative) helicity $k>0$ ($k<0$), whereas for negative energy values the pseudo-spin states are interchange themselves $v_{\uparrow}\to v_{\downarrow}$, where 
  the normalized  states $\{v_{\uparrow}, v_{\downarrow}\}$  are given by 
\begin{eqnarray}
v_{\uparrow}=\frac{1}{\sqrt{2}}\left(\begin{array}{c}1\\ 1\end{array}\right),~~~~~{\rm and}~~~~~v_{\downarrow}=\frac{1}{\sqrt{2}}\left(\begin{array}{c}1\\ -1\end{array}\right).
\end{eqnarray}
 For positive energy values one has the following two independent solutions $\Phi_{+, +}(x, \theta, t)=e^{ik x-i\frac{\left|E\right|t}{\hbar}}f(\theta)v_{\uparrow}$ and $\Phi_{+, -}(x, \theta, t)=e^{ik x-i\frac{\left|E\right|t}{\hbar}}f(\theta)v_{\downarrow}$  for positive and negative helicity, $k>0$ and $k<0$, respectively. In a similar fashion, for negative values of energy one has  $\Phi_{-, +}(x, \theta, t)=e^{ik x+i\frac{\left|E\right|t}{\hbar}}f(\theta)v_{\downarrow}$ and $\Phi_{-, -}(x, \theta, t)=e^{ik x+i\frac{\left|E\right|t}{\hbar}}f(\theta)v_{\uparrow}$  for positive and negative helicity, $k>0$ and $k<0$, respectively. These four states can be cast together as follows
 \begin{eqnarray}
 \Phi_{\mu, \sigma}(x, \theta, t)=e^{i\sigma \left|k\right| x-i\mu\frac{\left|E\right|t}{\hbar}}f(\theta)v_{\mu\cdot\sigma},
 \end{eqnarray}
 where we have introduced a mnemonic rule for $\mu\cdot\sigma$ as $+\cdot+=\uparrow$, $-\cdot+=\downarrow$, $+\cdot -=\downarrow$ and $-\cdot-=\uparrow$, where $\mu={\rm sgn} \left(E\right)$. Finally, let us recall that the transformation perfomed above, $\Psi=\Phi/\sqrt{r \Lambda}$, thus the asymptotic Dirac state faraway from scattered center is given by 
\begin{eqnarray}
 \Psi_{\mu, \sigma}(x, \theta, t)=\sqrt{V(x)}e^{i\sigma \left|k\right| x-i\mu\frac{\left|E\right|t}{\hbar}}f(\theta)v_{\mu\cdot\sigma},
 \label{instatesgen}
 \end{eqnarray}
which vanishes for $x\to\infty$. 
In addition, as a consequence of the periodicity of the function $f(\theta)=f(\theta+2\pi)$ one can write the following series representation $f(\theta)=\sum_{m\in \mathbb{Z}}f_{m}e^{i m\theta}$.

\section{Scattering analogue on the Bour's minimal surfaces}\label{sectV}

\subsection{Out-scattering states by Lippmann-Schwinger equation}

In this section, we introduce the Lippmann-Schwinger (LS) equation \cite{PhysRev.79.469, Sakurai1993Modern} in order to study how the states propagate along the the surface considered. In particular, we are interested to describe the manner how the initial states, found above, are scattered due to the effective potential $V(x)$. Let us consider the Hamiltonian $\hat{H}=\hat{H}_{0}+\hat{V}$  split between a ``free Hamiltonian" $\hat{H}_{0}$ and a perturbed potential $\hat{V}$. Now the LS equation is given by 
\begin{eqnarray}
\left|\Phi\right>=\left|\Phi_{\rm in}\right>+\frac{1}{E-\hat{H}_{0}+i\epsilon}\hat{V}\left|\Phi\right>,\label{LS}
\end{eqnarray}
where $\left|\Phi_{\rm in}\right>$ is the initial state, and $\left|\Phi\right>$ is the out scattering state. Note that in general we are considering that  $\hat{H}_{0}$ and $\hat{V}$ are differential matrix operators acting on spinors. Thus the states $\{\left|\Phi\right>\}$ acquire spinorial components. 

The Born approximation is gotten by substituting $\left|\Phi_{in}\right>$ instead of $\left|\Phi\right>$ in the second term of the LS equation (\ref{LS}). To go further to higher-order approximation it is standard to introduce the transition operator $\hat{T}$ 
defined using the  equation $\hat{T}\left|\Phi_{\rm in}\right>=\hat{V}\left|\Phi\right>$. In fact, multiplying  LS equation by $\hat{V}$ one arrives to the well-known self-consistent recursive operator equation for the transition operator $\hat{T}$ 
\begin{eqnarray}
\hat{T}=\hat{V}+\hat{V}\frac{1}{E-\hat{H}_{0}+i\epsilon}\hat{T}. 
\end{eqnarray}
A series solution for $\hat{T}$ can be gotten using this equation through a usual iterative procedure  
\begin{eqnarray}
\hat{T}&=&\hat{V}+\hat{V}\frac{1}{E-\hat{H}_{0}+i\epsilon}\hat{V}\nonumber\\
&+&\hat{V}\frac{1}{E-\hat{H}_{0}+i\epsilon}\hat{V}\frac{1}{E-\hat{H}_{0}+i\epsilon}\hat{V}+\cdots, \label{opT}
\end{eqnarray}
where the first approximation $\hat{T}\simeq \hat{V}$ corresponds to the so-called Born approximation. 

In order to determine the out scattering states one can follow two common procedures that are to project $\{\left|\Phi\right>\}$ along space states $\{\left|{\bf x}\right>\}$ or momentum states $\{\left|{\bf p}\right>\}$. For the projection along the space states one obtain the spinorial  wave function $\Psi({\bf x})=\left<{\bf x}\right|\left. \Phi\right>$, where ${\bf x}=(x_{1}, x_{2})$ are certain coordinates associated to a local patch on the surface.  Thus, using the transition operator the LS equation  can be re-written as 
\begin{eqnarray}
\Phi\left({\bf x}\right)=\Phi_{{\rm in}}\left({\bf x}\right)+\int_{\mathcal{D}}d^{2}{\bf x}^{\prime}~\mathbb{G}({\bf x}, {\bf x}^{\prime})\left<{\bf x}^{\prime}\right|\hat{T}\left|\Phi_{\rm in}\right>,\label{LS2}
\end{eqnarray}
where  the Green function satisfies the Green equation defined by 
\begin{eqnarray}
\left(E-\hat{H}_{0}\right)\mathbb{G}({\bf x}, {\bf x}^{\prime})=\mathbb{1}\delta\left({\bf x}-{\bf x}^{\prime}\right).
\end{eqnarray}
Now, when we project the out scattering states along the momentum states one obtain the spinorial wave function in the momentum space which we abusedly written with the same notation  $\Phi({\bf p})=\left<{\bf p}\right|\left. \Phi\right>$. Assuming that $\hat{H}_{0}$ is an operator that depends exclusively on the momentum operator, thus the LS equation can be written as 
\begin{eqnarray}
\Phi\left({\bf p}\right)=\Phi_{{\rm in}}\left({\bf p}\right)+\frac{1}{E-{H}_{0}({\bf p})+i\epsilon}
\left<{\bf p}\right|\hat{T}\left|\Phi_{\rm in}\right>\label{LS3},
\end{eqnarray}
where $H_{0}({\bf p)}$ is the matrix free Hamiltonian evaluated at the momentum value ${\bf p}$. 

\subsection{Out scattering  states on the Bour surfaces}

In this section, we study the electronic states on the Bour surfaces starting from the equation (\ref{DiracEqBour}). Noticeably, the second term of this equation can be thought of as a barrier potential. This  ``potential energy" is crucial in the behaviour of the Dirac particle states on the Bour surface. Now, in order to implement the Lippmann-Schwinger equation (\ref{LS}) we identify the free Hamiltonian $\hat{H}_{0}=v_{F}\sigma_{1}\hat{p}_{x}$ and the perturbed hamiltonian by $\hat{V}=v_{f}\sigma_{2}U(x)\hat{\ell}_{\theta}$ from (\ref{DiracEqBour}). In the following, we carry out the Born and higher-order Born approximations to determine the out scattering states on the curved surface.

\subsubsection{Born approximation.}
Now, in order to determine the out scattering  states $\Psi({\bf x})$, it is considered the initial states $\Phi_{in,\sigma}(x)$ faraway from the scattering center, which in our case corresponds to $x\simeq 0$. In particular, the states that are considered here are $\Phi_{in}({\bf x})=\Phi_{\mu, \sigma}(x, \theta, t)$  found  in last section (see states given by (\ref{instatesgen})).  Since the free Hamiltonian $\hat{H}_{0}$ is independent of $\theta$, thus the Green function acquires the expression $\mathbb{G}({\bf x}, {\bf x}^{\prime}; E)=\mathbb{G}(x-x^{\prime}; E)\delta(\theta-\theta^{\prime})$, where ${\bf x}=(x,\theta)$. Now,  the Lippmann-Schwinger equation in this case can be written as
\begin{eqnarray}
\Phi(x, \theta)&=&\Phi_{\pm}(x, \theta)\nonumber\\
&+&v_{F}\int_{-\infty}^{\infty}dx^{\prime}~ \mathbb{G}(x- x^{\prime};  E)U(x^{\prime})\sigma_{2}\hat{\ell}_{\theta}\Phi(x^{\prime},\theta),\nonumber\\
\end{eqnarray}
where $\mathbb{G}(x-x^{\prime}; E)$ is the  Green function for the one-dimensional operator $E-v_{F}\sigma_{1}\hat{p}_{x}$, that is, a function that satisfies  $\left(E-v_{F}\sigma_{1}\hat{p}_{x}\right)\mathbb{G}(x-x^{\prime}; E)=\delta(x-x^{\prime})\mathbb{1}$. Following, the standard procedure  (see appendix \ref{AppA}) it is not difficult to show that 
{\small\begin{eqnarray}
\mathbb{G}(x-x^{\prime}; E)=\frac{1}{2i \hbar v_{F}}\left[{\rm sgn}(E)+{\rm sgn}(x-x^{\prime})\sigma_{1}\right]e^{ i\frac{\left|E\right|}{\hbar v_{F}}\left|x-x^{\prime}\right|},\nonumber\\ \label{GFcoord}
\end{eqnarray}}
where we recall that $E=\pm \hbar v_{F}\left|k\right|$. In the following, as a consequence of the polar geometry of the surface, the states  $\Phi(x, \theta)$ are periodic in the angular variable $\theta$, thus it can be written in the next expansion
$\Phi(x, \theta)=\sum_{m\in\mathbb{Z}}\Phi_{m}(x)e^{im\theta}$.
Now, using the orthonormal relation of the basis $\{e^{im\theta}\}$ one have the following integral equation for $\Phi^{(\mu)}_{m}(x)$, that is, 
\begin{eqnarray}
\Phi^{(\mu)}_{m}(x)&=&e^{i\sigma \left|k\right| x}f_{m}v_{\mu\cdot \sigma}\nonumber\\ &+&\hbar v_{F}m\int_{-\infty}^{\infty}dx^{\prime}\mathbb{G}(x- x^{\prime}; E)U(x^{\prime})\sigma_{2}\Phi^{(\mu)}_{m}(x^{\prime}),\nonumber\\
\label{integraleq1}
\end{eqnarray}
where we have introduced the labeled $\mu$ in order to distinguish the positive and energy out scattering states; recall also that $\mu={\rm sgn}(E)$.  At the Born approximation it is enough to make the substitution $\Phi^{(\mu)}_{m}(x^{\prime})$ by $e^{i\sigma p x^{\prime}}f_{m}v_{\mu\cdot \sigma}$  in the second term of last equation, where we have defined the magnitude of the momentum $\left|k\right|=p$. Let us choose a initial wave with $\sigma=-$, that is, a left wave  with $k<0$ going to the scattering center. 
Noticeably, after a straightforward calculation the states are given by
\begin{eqnarray}
\Phi^{(\mu)}_{m}(x)&\simeq&  f_{m}\left[e^{-ip x}v_{-\mu}+m ~\tilde{U}\left(2p \right)e^{ipx}v_{\mu}\right],\label{BornApproximation}
\end{eqnarray}
where we have found that negative energy states with  pseudo-spin up $\uparrow$ propagating along the scattering center reflecting into a pseudospin down $\downarrow$. The amplitud of the reflection is given by $m ~\tilde{U}(2p)$, where $\tilde{U}$ is the Fourier transform of the effective potential $U(x)$. In order to have a better understanding of this scattering phenomena we proceed to carry out a higher-order Born approximation in the following section.

\subsubsection{Higher-Order Born approximation.}

For the higher-order Born approximation, we found more useful the momentum representation of the states. The starting point is the equation for the out scattering states (\ref{LS3}), where the term $\left<{\bf p}\right|\hat{T}\left|\Phi_{\rm in}\right>$ is determined approximately by using the series approximation of the recursive equation for the transition operator (\ref{opT}), that is, 
\begin{eqnarray}
\left<{\bf p}\right|\hat{T}\left|\Phi_{\rm in}\right>&=&\left<{\bf p}\right|\hat{V}\left|\Phi_{\rm in}\right>+\left<{\bf p}\right|\hat{V}\hat{G}\hat{V}\left|\Phi_{\rm in}\right>+\cdots\nonumber\\&+&\left<{\bf p}\right|\hat{V}\hat{G}\hat{V}\cdots \hat{V}\hat{G}\hat{V}\left|\Phi_{\rm in}\right>+\cdots,
\label{Tseries}
\end{eqnarray}
where we have defined the resolvent operator $\hat{G}:=1/(E-\hat{H}_{0}+i\epsilon)$. Now, for each term of $\hat{G}$  of last expansion one introduce two completeness relation in momentum space $\mathbb{1}=\sum_{{\bf q}}\left|{\bf q}\right>\left<{\bf q}\right|$, where $\sum_{{\bf q}}=\frac{1}{2\pi}\sum_{m}\int \frac{dq}{2\pi}$ and  $\left|{\bf q}\right>:=\left|q, m\right>$. These completeness relations are introduced before and after the operator $\hat{G}$. Thus, one has the first term $\tau_{1}(p):=\left<{\bf p}\right|\hat{V}\left|\Phi_{\rm in}\right>$ and 
the $(n+1)-{\rm th}$ term, with $n\geq 1$, has the following structure  
\begin{eqnarray}
\boldsymbol{\tau}_{n+1}\left(p\right):&=&\sum_{{\bf q}^{(1)}, \cdots, {\bf q}^{(2n)}}\left<{\bf p}\right|\hat{V}\left|{\bf q}^{(1)}\right>\left<{\bf q}^{(1)}\right|\hat{G}\left|{\bf q}^{(2)}\right>\nonumber\\&\times&\left<{\bf q}^{(2)}\right|\hat{V}\cdots \hat{V}\left|{\bf q}^{(2n-1)}\right>\left<{\bf q}^{(2n-1)}\right|\hat{G}\left|{\bf q}^{(2n)}\right>\nonumber\\&\times& \left<{\bf q}^{(2n)}\right|\hat{V}\left|\Phi_{\rm in}\right>.
\end{eqnarray}

Now, one can reduce half of the integrals since for each term $\left<{\bf q}^{(i)}\right|\hat{G}\left|{\bf q}^{(j)}\right>=\mathbb{G}(q^{j})\delta_{{\bf q}_{i},{\bf q}_{j}}$, where $\mathbb{G}({\bf q}^{j})=1/(E-H_{0}({\bf q})+i\epsilon)$. Thus the  $(n+1)-{\rm th}$ can be simplify as follows

\begin{widetext}
\begin{eqnarray}
\boldsymbol{\tau}_{n+1}\left(p\right)&=&{\sum_{{\bf q}^{(1)}, \cdots, {\bf q}^{(n)}}\left<{\bf p}\right|\hat{V}\left|{\bf q}^{(1)}\right>\left(\prod_{\ell=1}^{n-1}\mathbb{G}({\bf q}^{\ell})\left<{\bf q}^{(\ell)}\right|\hat{V}\left|{\bf q}^{(\ell+1)}\right>\right)}\mathbb{G}({\bf q}^{(n)})\left<{\bf q}^{(n)}\right|\hat{V}\left|\Phi_{\rm in}\right>,\label{termkm1}
\end{eqnarray}
\end{widetext}

where the state $\left|{\bf q}^{\ell}\right>=\left|q^{\ell}, m^{\ell}\right>$, for $\ell=1,2,\cdots, n$. 
In order to simplify last expression  it is necessary to find the following two generic expressions  ({\it a.}) $\left<{\bf p}\right|\hat{V}\left|{\bf q}\right>$, and  ({\it b.}) $\left<{\bf p}\right|\hat{V}\left|\Phi_{\rm in}\right>$. The momentum states are given expressed by $\left|{\bf p}\right>=\left|p, m\right>$ and similarly $\left|{\bf q}\right>=\left|q, m^{\prime}\right>$. Thus for the terms type ({\it a.}) can be written as $\left<{\bf p}\right|\hat{V}\left|{\bf q}\right>=(2\pi)\hbar v_{F}\sigma_{2}m\delta_{mm^{\prime}}\left<p\right|\hat{U}(x)\left|q\right>$ where we have acted the angular momentum operator $\hat{\ell}_{\theta}\left|m\right>=\hbar m\left|m\right>$ and introduced the orthogonal relation $\left<m\right|\left. m^{\prime}\right>=2\pi \delta_{mm^{\prime}}$.  Now, we introduce the completeness relation $\mathbb{1}=\int dx\left|x\right>\left<x\right|$ and we use $\left<x\right|\left. p\right>=e^{i p x}$, thus one has 
\begin{eqnarray}
\left<{\bf p}\right|\hat{V}\left|{\bf q}\right>=(2\pi)\hbar v_{F}\sigma_{2}m\delta_{mm^{\prime}}\tilde{U}\left(p-q\right)\label{typea}
\end{eqnarray}
being $\tilde{U}(q):=\int dx e^{-iqx}U(x)$ the Fourier transform of the potential. Now, for the type ({\it b}.) terms one introduce a completeness relation in the momentum space such that  $\left<{\bf p}\right|\hat{V}\left|\Phi_{\rm in}\right>=\sum_{{\bf q}}\left<{\bf p}\right|\hat{V}\left|{\bf q}\right>\left<{\bf q}\right.\left|\Phi_{\rm in}\right>$. Now, it is necessary to calculate the term $\Psi_{in}({\bf p})=\left<{\bf q}\right.\left|\Phi_{\rm in}\right>$, which is the Fourier transform of the initial wave $e^{i\sigma \left|k\right| x}f_{m}v_{\mu\cdot \sigma}$ that is $\Psi_{in}({\bf p})=2\pi \delta\left(p-\sigma\left|k\right|\right)f_{m}v_{\mu\cdot \sigma}$, thus one has  
\begin{eqnarray}
\left<{\bf p}\right|\hat{V}\left|\Phi_{\rm in}\right>
=\hbar v_{F} \sigma_{2}v_{\mu\cdot\sigma}f_{m}m\tilde{U}\left(p-\sigma\left|k\right|\right).\label{typeb}
\end{eqnarray}
Notice that r.h.s. of last equation corresponds to the first term in the series (\ref{Tseries}), that is,  
\begin{eqnarray}
\tau_{1}(p)=\hbar v_{F} f_{m} \left(i\mu m\right) \tilde{U}\left(p+\left|k\right|\right) v_{\mu},
\end{eqnarray}
where we have put $\sigma=-1$ since we have an initial left wave, and where we have used the identity $\sigma_{2}v_{-\mu}=i\mu v_{\mu}$.  Now, we introduce the terms type ({\it a.}) (\ref{typea}) and ({\it b.}) (\ref{typeb}) in the ($n+1$-th) term $\boldsymbol{\tau}_{n+1}\left(p\right)$ (\ref{termkm1}). Now, again one has to put $\sigma=-1$, and following the straightforward caculation developed in the appendix (\ref{AppB}) one is able to find 
  {\small\begin{eqnarray}
\tau_{n+1}=\hbar v_{F}f_{m}\left(i\mu  m\right)^{n}m\tilde{U}\left(p-\left|k\right|\right)\tilde{U}\left(2\left|k\right|\right)\left|\tilde{U}\left(2\left|k\right|\right)\right|^{n-1}v_{-\mu},\nonumber\\\label{tauoddn}
\end{eqnarray}}
for odd $n$, whereas 
\begin{eqnarray}
\tau_{n+1}=\hbar v_{F}f_{m}\left(i\mu  m\right)^{n+1}\tilde{U}\left(p+\left|k\right|\right)\left|\tilde{U}\left(2\left|k\right|\right)\right|^{n}v_{\mu},\nonumber\\
\label{tauevenn}
\end{eqnarray}
for even $n$, where $\left|\cdot\right|$ is the complex norm. In this manner the series (\ref{Tseries}) is $\left<{\bf p}\right|\hat{T}\left|\Phi_{\rm in}\right>=\sum_{n=0}^{\infty}\tau_{n+1}\left(p\right)$. This expectation value must be introduce in the LS equation (\ref{LS3}). Afterwards, one need to compute the Fourier transform in order to find an expression of the Dirac wave 
\begin{eqnarray}
\Phi^{(\mu)}_{m}(x)=e^{-i \left|k\right| x}f_{m}v_{-\mu}+\sum_{n=0}^{\infty}C_{n+1}(x),\label{wave1}
\end{eqnarray}
where one has still to compute the inverse Fourier transform $C_{n+1}(x)=\int \frac{dp}{2\pi}e^{ipx}\mathbb{G}(p)\tau_{n+1}\left(p\right)$. See appendix (\ref{AppB}) for specific details calculation of these integrals. 
Thus the result for $C_{n+1}(x)$ is the following, for odd integers $n=2j+1$
\begin{eqnarray}
C_{2j+2}(x)=\left(-1\right)^{j+1}\left(m^2\left|\tilde{U}\left(2\left|k\right|\right)\right|^2\right)^{j+1}f_{m}e^{-i\left|k\right| x}v_{-\mu}\nonumber\\\label{Coddj}
\end{eqnarray}
while for even integers $n=2j$
{\small\begin{eqnarray}
C_{2j+1}(x)=\left(-1\right)^{j+1}m\tilde{U}\left(2\left|k\right|\right)\left(m^2\left|\tilde{U}\left(2\left|k\right|\right)\right|^2\right)^{j}f_{m}e^{i\left|k\right| x}v_{\mu}\nonumber\\
\label{Cevenj}
\end{eqnarray}}
for $j\in \mathbb{N}\cup \{0\}$. Now, after insert these expressions in (\ref{wave1}) it is noticeably each of the factors appearing in front of the ongoing ($e^{-i\left|k\right|x}$) and incoming  ($e^{i\left|k\right|x}$) terms can be cast as a geometric series that can be sum up as $\sum_{\ell=0}^{\infty}(-1)^{\ell}a^{\ell}\left(m,k\right)=1/(1+a\left(m, k\right))$ while $a\left(m, k\right)<1$, where $a\left(m, k\right)=m^2 \left|\tilde{U}\left(2\left|k\right|\right)\right|^{2}$. Thus the final expression for the  Dirac wave is 
\begin{eqnarray}\Phi^{(\mu)}_{m}(x)=f_{m}\left[\mathcal{F}(m, k)e^{-i\left|k\right|x}v_{-\mu}+\mathcal{G}(m, k)e^{i\left|k\right|x}v_{\mu}\right]\nonumber\\
\label{resultcompleto}
\end{eqnarray}
where the coefficients $\mathcal{F}(m, k)$ and $\mathcal{G}(m, k)$ are given by 
\begin{eqnarray}
\mathcal{F}(m, k)&=&\frac{1}{1+m^{2}\left|\tilde{U}\left(2\left|k\right|\right)\right|^2},\label{Feq}\\
\mathcal{G}(m, k)&=& m\tilde{U}\left(2\left|k\right|\right)\mathcal{F}(m, k).
\label{GEq}
\end{eqnarray}
The Born approximation (\ref{BornApproximation}) is recovered when $\mathcal{F}\left(m, k\right)\approx 1$. This is expected to be achivied for large values of momenta $\left|k\right|$.

Since one can prepare a initial Dirac wave with values of $m$ and $k$ such that $a\left(m, k\right)\geq 1$, thus here we consider an analytical continuation of the geometric series to take into account values of $m$ and $k$ under last condition. This analytical continuation means that the factors $\mathcal{F}(m, k)$ and $\mathcal{G}(m, k)$ have the same function for values of $(m, k)$ in region where $a\left(m, k\right)\geq 1$.

\section{ Dirac waves transmission through the Bour surfaces, and the Klein paradox}\label{sectVI}

\begin{figure}[ht]
    \centering
   \includegraphics[scale=0.17]{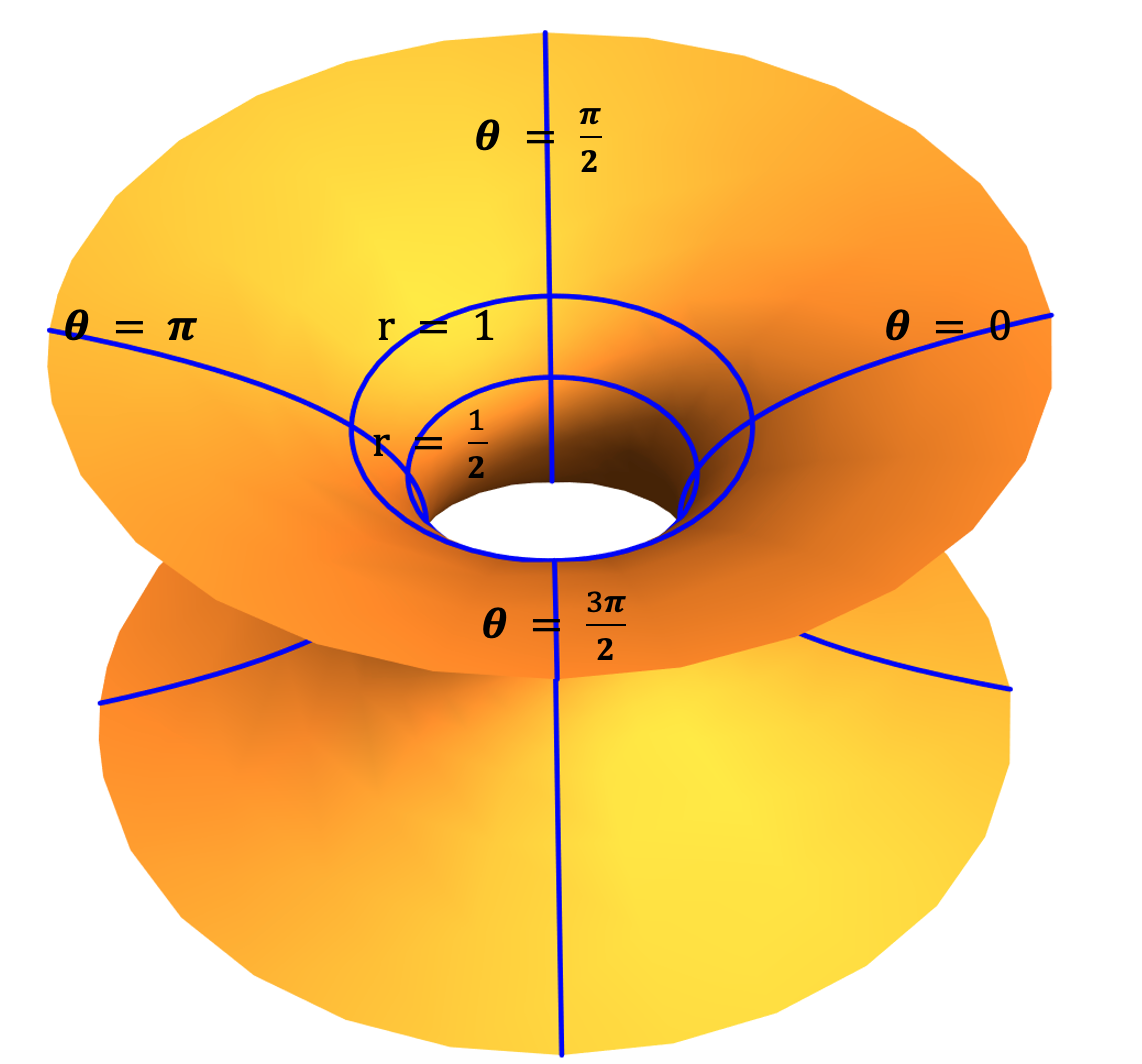}\\
   \includegraphics[scale=0.17]{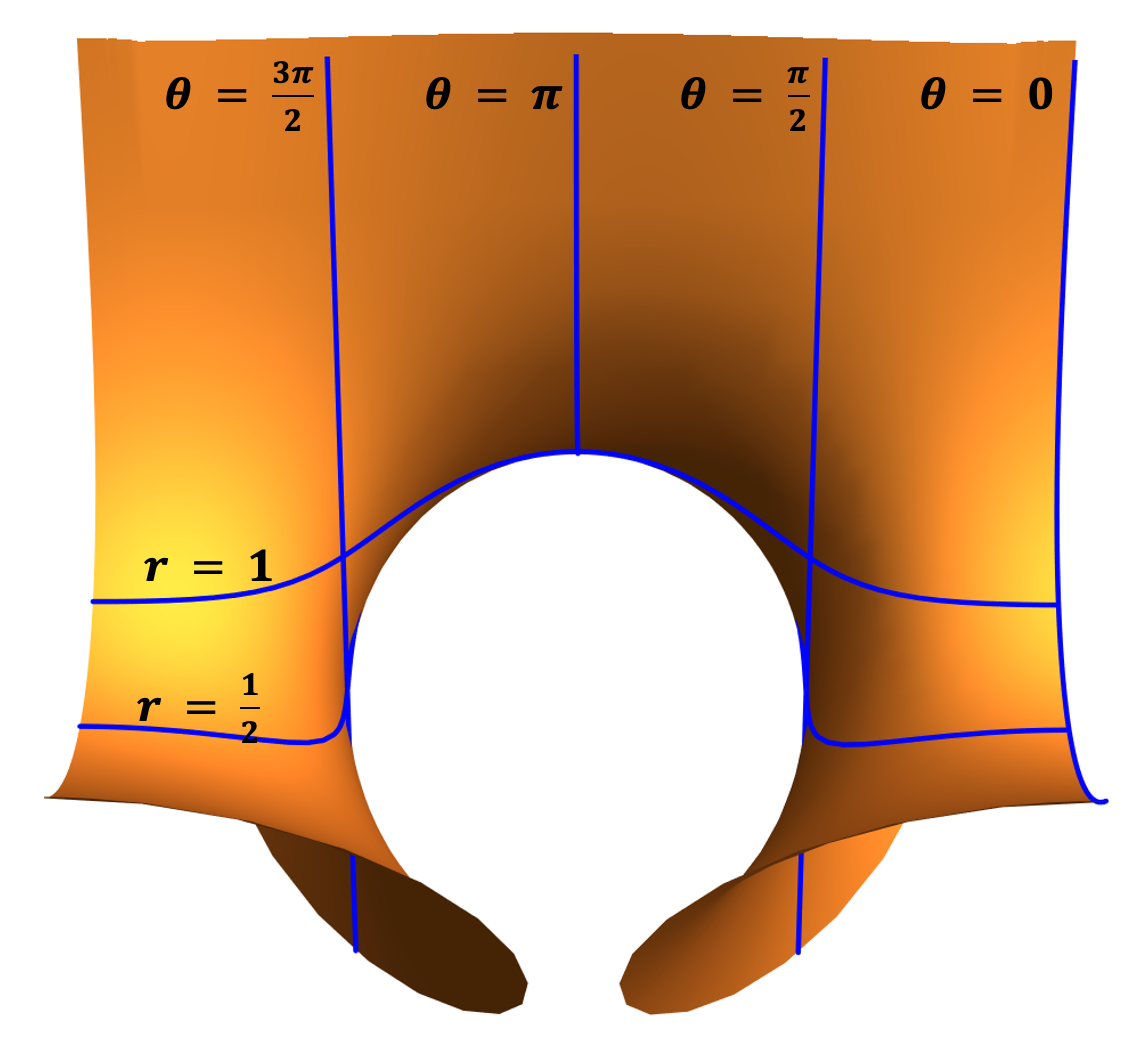}\\
   \includegraphics[scale=0.17]{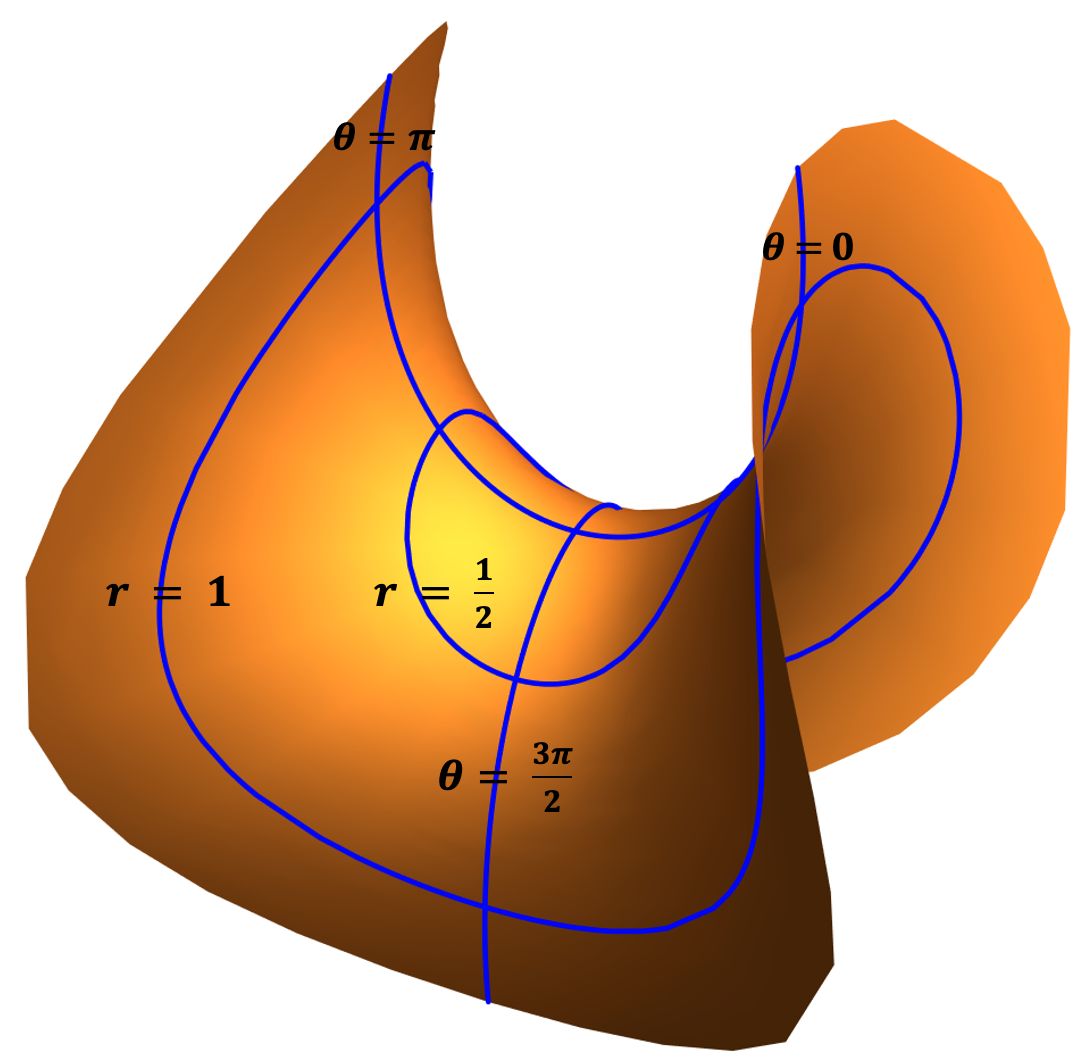}
    \caption{ Catenoid, $B_{1}$-Bour and classical Enneper minimal surface, from the top to the bottom, drawn with the parametrizations (\ref{ParamCat}), (\ref{ParamB1}) and (\ref{ParamEnneper}). Each of the surface include examples of level curves with $r$ and $\theta-$constant .}
    \label{fig223}
\end{figure}

In this section, we introduce the N$\ddot{\rm o}$ether current $J^{\mu}$ of the Dirac equation as the probability current density. This quantity is introduced in order to study the manner how the initial Dirac wave is propagated along the surface. In particular, using the current $J^{\mu}$  we are able to determine the transmission and reflection coefficients.  This conserved quantity is given by \cite{parker_toms_2009}
\begin{eqnarray}
J^{\mu}=\overline{\Psi}\underline{\gamma}^{\mu}\Psi,\label{NoetherCurrent}
\end{eqnarray}
where  $\overline{\Psi}=\Psi^{\dagger}\underline{\gamma}^{0}$. Also, we recall that $\underline{\gamma}^{\mu}$ are given in terms of the vielbeins and the Dirac matrices $\gamma^{A}$ introduced above in section ({\ref{DiracEq}}). Also, recall that $\Psi(x)=r^{-1/2}\Lambda^{-1/2}\Phi$, where $\Lambda$ is the conformal factor introduced above. The zero component of the current, $J^{0}(x)$, allows us to determine the probability density function, whereas the spatial components of the current allows us to determine how the Dirac wave propagate through the space geometry, that is, using the spatial components one can define the reflection coefficient as 
\begin{eqnarray}
\mathcal{R}=\left|\frac{J^{a}_{\rm ref}n_{a}}{J^{b}_{\rm inc}n_{b}}\right|\label{defR}
\end{eqnarray}
and the transmission coefficient by its complement $\mathcal{T}=1-\mathcal{R}$, where ${\bf n}$ is a tangent vector on the surface that is normal to a curve $\gamma$ embedded on the surface. In particular, for the Bour surface we choose $\gamma$ to be a  $r-$constant curve, thus ${\bf n}$ is a tangent vector along the $\theta$ direction. See figure (\ref{fig223}) to see level curves with $r-$constant and $\theta-$constant on the catenoid, $B_{1}$ and Enneper surfaces. 

In the following, we focus to determine the general expressions for the probability density $J^{0}$, transmission coefficient $\mathcal{T}$, reflection coefficient $\mathcal{R}$ and conductance $G$ for the Bour surfaces, where particular emphasis is made on the catenoid/helicoid, $B_{1}$-Bour and classical Enneper surface.

\subsection{Probability density function $J^{0}$}
Before to present the result for the probability density function let us obtain  normalization factors $\mathcal{N}_{0}$, for the free wave (\ref{instatesgen}), and $\mathcal{N}$, for the scattered wave (\ref{resultcompleto}). For this purpose, consider a large portion of area of the surface, and let us impose the condition $\int_{\mathcal{D}}dA J^{0}=1$, where $dA=r \Lambda(r)drd\theta$ is the area element in the surface. This condition  guarantees that a Dirac fermion is surely at some point, $(r, \theta)\in \mathcal{D}=[0, L]\times[0, 2\pi]$, from the domain $\mathcal{D}$. 

Now, the zero component of the N$\ddot{\rm o}$ether current is the probability density function given for our particular space-time geometry (see (\ref{SpaceTimePolar})) by 
$J^{0}(r)=\frac{1}{v_{F}}\Psi^{\dagger(\mu)}_{m}\Psi^{(\mu)}_{m}$. Thus for the free initial wave one has the density  $J^{0}(r)=\frac{\mathcal{N}^2_{0}\left|f_{m}\right|^2}{v_{F}r\Lambda}$, thus the normalization factor is easily obtained as
$\mathcal{N}_{0}=\left(v_{F}/(2\pi L)\right)^{\frac{1}{2}}/\left|f_{m}\right|$, while for the scattered wave $J^{0}(r)=\frac{\mathcal{N}^2\left|f_{m}\right|^2}{v_{F}r\Lambda}\mathcal{F}\left(m, k\right)$ the normalization factor is given by 
$ \mathcal{N}=\mathcal{N}_{0}/\sqrt{\mathcal{F}\left(m, k\right)}$.  Now, the probability density function is given  for both waves the probability density by the expression 
\begin{eqnarray}
J^{0}(r,\theta)=\frac{1}{2\pi L r \Lambda\left(r\right)},
\end{eqnarray}
which means that it is most probable to find Dirac particles near the scattering center on the Bour surface. It is noteworthy to mention that the scattering point ($x\simeq 0$)  corresponds to the point where the curvature attains its maximum value (see figure (\ref{fig1})).

\subsection{Transmittance and conductance on the Bour geometries.} 
Now, we want to determine the reflection and the transmission coefficients. We choose $\gamma$ to be the $r-$constant curve on a Bour surface, thus the normal vector to $\gamma$ is tangent to a  $\theta$-constant curve. Notice that $J^{\theta}=-\frac{1}{r^2\Lambda^2}\Phi^{\dagger(\mu)}_{m}\sigma_{1}\Phi^{(\mu)}_{m}$ after using $\underline{\gamma}^{\theta}=\gamma^{2}e^{\theta}_{1}$, where the vielbein in this case is $e^{\theta}_{1}=1/\Lambda$. Now, recall that $\sigma_{1}v_{\mu}=\mu v_{\mu}$ and $v_{\mu}^{\dagger}v_{\mu}=1$, we compute the incidence current $J^{\theta}_{\rm inc}$ using the initial wave $\mathcal{N}_{0}e^{-i\left|k\right|x}f_{m}v_{-\mu}$ 
\begin{eqnarray}
J^{\theta}_{\rm inc}=\frac{v_{F} \mu}{(2\pi L)r^2\Lambda^2},
\end{eqnarray}
and  compute $J_{\rm ref}^{\theta}$  using the reflection wave $\mathcal{N}f_{m}\mathcal{G}(m, k)e^{i\left|k\right|x}v_{\mu}$, getting
\begin{eqnarray}
J_{\rm ref}^{\theta}=-\frac{v_{F} \mu}{(2\pi L)r^2\Lambda^2}\frac{\left|\mathcal{G}\left(m, k\right)\right|^{2}}{\mathcal{F}\left(m, k\right)}.
\end{eqnarray}
Now, using the reflection and transmission coefficient definition (\ref{defR}), it is not difficult to get  $\mathcal{R}=m^{2}\tilde{U}^2\left(2\left|k\right|\right)\mathcal{F}(m, k)$ and $\mathcal{T}=\mathcal{F}\left(m, k\right)$, respectively, where $\mathcal{F}(m, k)$ is given by Eq. (\ref{Feq}).

Both coefficients $\mathcal{R}$ and $\mathcal{T}$ are given in terms of the Fourier transform  $\tilde{U}(p)=\int_{-\infty}^{\infty}dx e^{-ip x}U(x)$, thus it is convenient to make further simplifications. Since the potential is proportional to $1/|c|$ it is convenient to perform the following change of variable $\tilde{x}=x/|c|$, thus let us define the function
\begin{eqnarray}
\mathcal{U}(\xi)=\int_{-\infty}^{\infty}d\tilde{x}e^{-i \xi \tilde{x}}\left|c\right|U(\left|c\right|\tilde{x}).
\end{eqnarray}
 Now, instead of using the wave number $k$ we use the energy dispersion relation $E=\pm \hbar v_{F}\left|k\right|$.  Also, notice that the Bour material introduces a natural scale of energy given by $E_{0}=\hbar v_{F}/\left|c\right|$ in terms of the characteristic length, $\left|c\right|$, associated to each Bour surface.  Therefore the reflection and transmission coefficient in terms of the energy $E$ are given for any Bour surface within the present approximation as 
 
\begin{eqnarray}
\mathcal{R}(E)&=&\frac{m^2\left|\mathcal{U}\left(2E_{*}\right)\right|^{2}}{1+m^2\left| ~\mathcal{U}\left(2E_{*}\right)\right|^2},\label{GenRes0}\\
\mathcal{T}(E)&=&\frac{1}{1+m^{2}\left|\mathcal{U}\left(2E_{*}\right)\right|^2},\label{GenRes}
\end{eqnarray}
where  $E_{*}=\left|E\right|/E_{0}$ is a dimensionless parameter. The conductance can be computed using the simple expression $G\left(E\right)=\frac{e^2}{\pi \hbar}\mathcal{T}\left(E\right)$. Notice that each Bour material with typical characteristic length  $\left|c\right|$ introduces a natural scale of energy  $E_{0}=\hbar v_{F}/\left|c\right|$; for surfaces with $\left|c\right|\sim 1~{\rm nm}$ the characteristic scale energy is $E_{0}\sim 8.27~{\rm eV}$. Note that for big values of energy, that is, $\left|E\right|\gg E_{0}$, the above result (\ref{GenRes0}) and (\ref{GenRes}) reduces to the Born approximation since in this parameter region one has $\mathcal{R}(E)\simeq m^2\left|\mathcal{U}\left(2E_{*}\right)\right|^{2} $ and  $\mathcal{T}(E)\simeq 1-m^2\left|\mathcal{U}\left(2E_{*}\right)\right|^{2}$, which is the Born approximation result. 

Now, we still need to compute the Fourier transform of the effective potential. Conspicuously, for any Bour surface it is useful to back to the original radial coordinate instead of $x$ since $dx= \Lambda(r) dr$ and the effective potential $V(r)=1/(r \Lambda \left(r\right))$, thus Fourier transform is 
\begin{eqnarray}
\mathcal{U}(\xi)=\int_{0}^{\infty}\frac{dr}{r}\exp\left[-i \xi \frac{x_{n}(r)}{\left|c\right|}\right],\label{FourierMU}
\end{eqnarray}
where for each  Bour surface labeled with $n$, $x_{n}(r)$ is the change of variable introduced above (see for instance Eq. (\ref{ChVar}) for $n\neq 1$.).

\subsubsection{Transmission on the catenoid/helicoid.} In this case one has $x_{0}(r)=\left|c\right|\left(r-r^{-1}\right)$. Here, it is convenient to make the change of variable $y=\log r\in \mathbb{R}$, thus the argument of the exponential in (\ref{FourierMU}) turns out to be the odd function  $\sinh\left(y\right)$, meaning  thas  Fourier transform is simplified to $\mathcal{U}(\xi)=\int_{-\infty}^{\infty}dy\cos\left[2\xi \sinh\left(y\right)\right]$. This integral can be expressed in terms of a modified Bessel function \cite{gradshteyn2007}
\begin{eqnarray}
\mathcal{U}(\xi)=2K_{0}(2\xi).\label{FourierCat}
\end{eqnarray}
 
\begin{figure}[ht]
    \centering
       \includegraphics[scale=0.3]{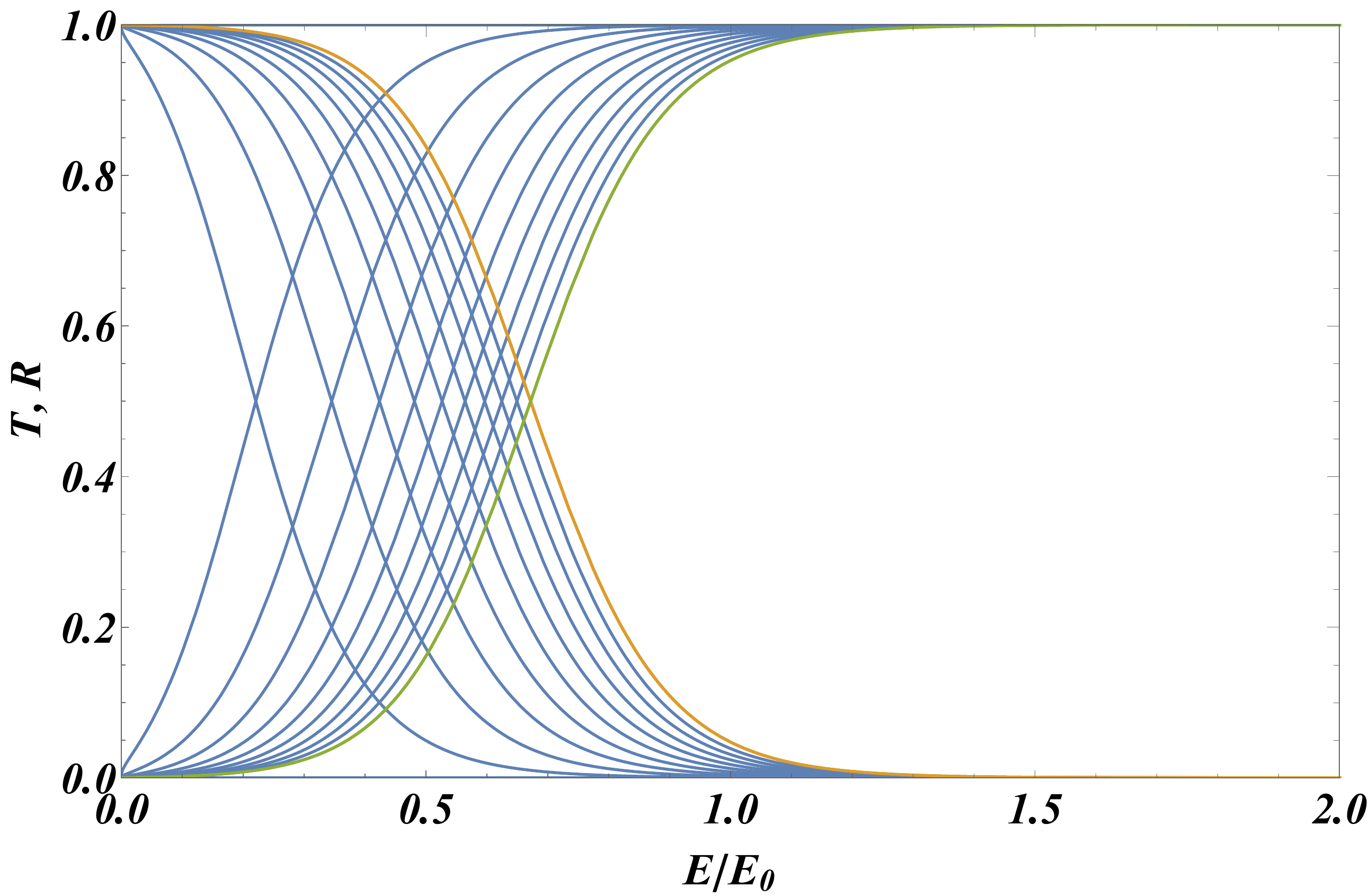}
    \caption{ Family of transmission and reflection coefficients (Eqs.(\ref{GenRes0}) and (\ref{GenRes}))  versus reduced energy $E_{*}=E/E_{0}$ using the Fourier transform (\ref{FourierCat}) corresponding to  the catenoidal/helicoidal geometry. The set of curves were obtained for cases with $m=1, \cdots, 10$. The orange and green curves are guides for the eyes to identify the $m=10$ case.}
    \label{RefTransCatenoide}
\end{figure}

In figure (\ref{RefTransCatenoide}), it is shown the reflection and transmission coefficient for a propagation wave through the catenoid, where we have been used (\ref{FourierCat}). In addition, one can see that the value of $E_{*}=E_{\rm x}$, where the transmission and reflection have the same value is translated to the right as long as the value of $m$ increases. The interception  value of energy is  $E_{\rm x}$, where the  transcendental equation is satisfied  
\begin{eqnarray}
K_{0}(4E_{\rm x})=\frac{1}{2m}.
\end{eqnarray}
 It is also noteworthy to mention, that at the interception the  value of the transmission coefficient (or the reflection coefficient) is independent of $m$ and given by  $\mathcal{T}(E_{\rm x})=1/2$, as can be appreciate in Fig.  (\ref{RefTransCatenoide}). 

\subsubsection{Transmission on $B_{1}$-Bour.} In this case one has $x_{1}(r)=\left|c\right|\left(\log r+\frac{1}{2}r^{2}\right)$. Here, let us substitute $i\xi\to z$, where $z$ is a complex value with $\Re z<0$. Additionally, it is convenient to make the change of variable $y=r^2\in \mathbb{R}^{+}$, thus the integral  (\ref{FourierMU}) turns out to be $\mathcal{U}\left(\xi\right)=2\int_{0}^{\infty}\frac{dy}{y} y^{-\frac{z}{2}}e^{-\frac{z}{2}y}$, which can be related to a Gamma function \cite{gradshteyn2007} as
\begin{eqnarray}
\mathcal{U}(\xi)=\lim_{z\to i\xi}2\left(\frac{z}{2}\right)^{\frac{z}{2}}\Gamma\left(-\frac{z}{2}\right).\label{FourierB1}
\end{eqnarray}
Since we need the complex norm of $\mathcal{U}(\xi)$, we use the identity $\left|\Gamma(i y)\right|^2=\pi/(y \sinh(\pi y))$ for $y\in\mathbb{R}$ \cite{gradshteyn2007}, it remains to compute the complex norm of the factor in (\ref{FourierB1}), by straightforward elementary calculation one has $\left(\frac{z}{2}\right)^{\frac{z}{2}}=e^{i\frac{\xi}{2}\log\frac{\xi}{2}}e^{-\frac{\xi\pi}{4}}$. Therefore 
\begin{eqnarray}
\left|\mathcal{U}(\xi)\right|^{2}=\frac{4}{\xi\left(e^{\pi \xi}-1\right)}.\label{FourierRealB1}
\end{eqnarray}

\begin{figure}[ht]
    \centering
       \includegraphics[scale=0.34]{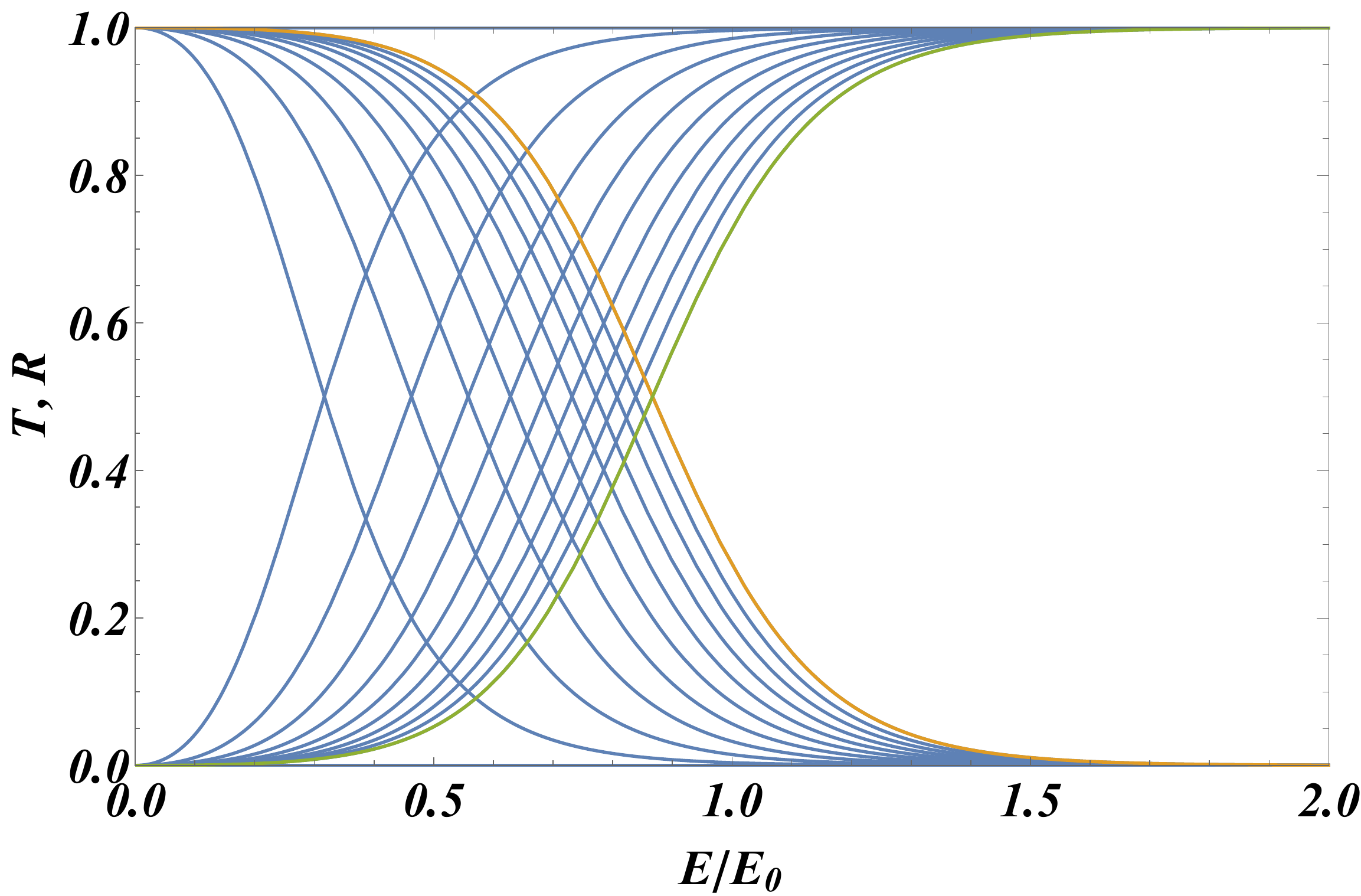}
    \caption{ Family of transmission and reflection coefficients (Eqs.(\ref{GenRes0}) and (\ref{GenRes}))  versus reduced energy $E_{*}=E/E_{0}$ using the Fourier transform (\ref{FourierB1}) corresponding to  the $B_{1}$ geometry. The set of curves were obtained for cases with $m=1, \cdots, 10$. The orange and green curves are guides for the eyes to identify the $m=10$ case.
}
    \label{RefTransB1}
\end{figure}

In figure (\ref{RefTransB1}), it is shown the reflection and transmission coefficient for a propagation wave through the $B_{1}$-Bour surface, where we have been used (\ref{FourierRealB1}). The structure of these curves is very similar to the previous case. The principal difference corresponds that the total transmission occurs to a bigger value of $E_{*}$. In addition, how the curves are moving to the left is given according to the transcendental equation
\begin{eqnarray}
E_{x}\left(e^{2\pi E_{x}}-1\right)=2m^{2}.
\end{eqnarray}

\subsubsection{Transmission on the classical Enneper surface and $n> 2$ Bour surfaces.} In this case one has $x_{n}(r)=\left|c\right|\left(\frac{r^{n-1}}{n-1}+\frac{r^{n+1}}{n+1}\right)$ with $n\geq 2$, thus the Fourier transform reduces to the integral
\begin{eqnarray}
\mathcal{U}\left(\xi\right)=2\int_{0}^{\infty}\frac{dr}{r} \cos\left(\xi \left(\frac{r^{n-1}}{n-1}+\frac{r^{n+1}}{n+1}\right)\right).\label{FourierTransform} 
\end{eqnarray}
This integral is strictly divergent due to the singularity at $r=0$. The integral is regularized introducing an inferior cut-off such that $r\geq \epsilon$. 

\begin{figure}[ht]
    \centering
       \includegraphics[scale=0.43]{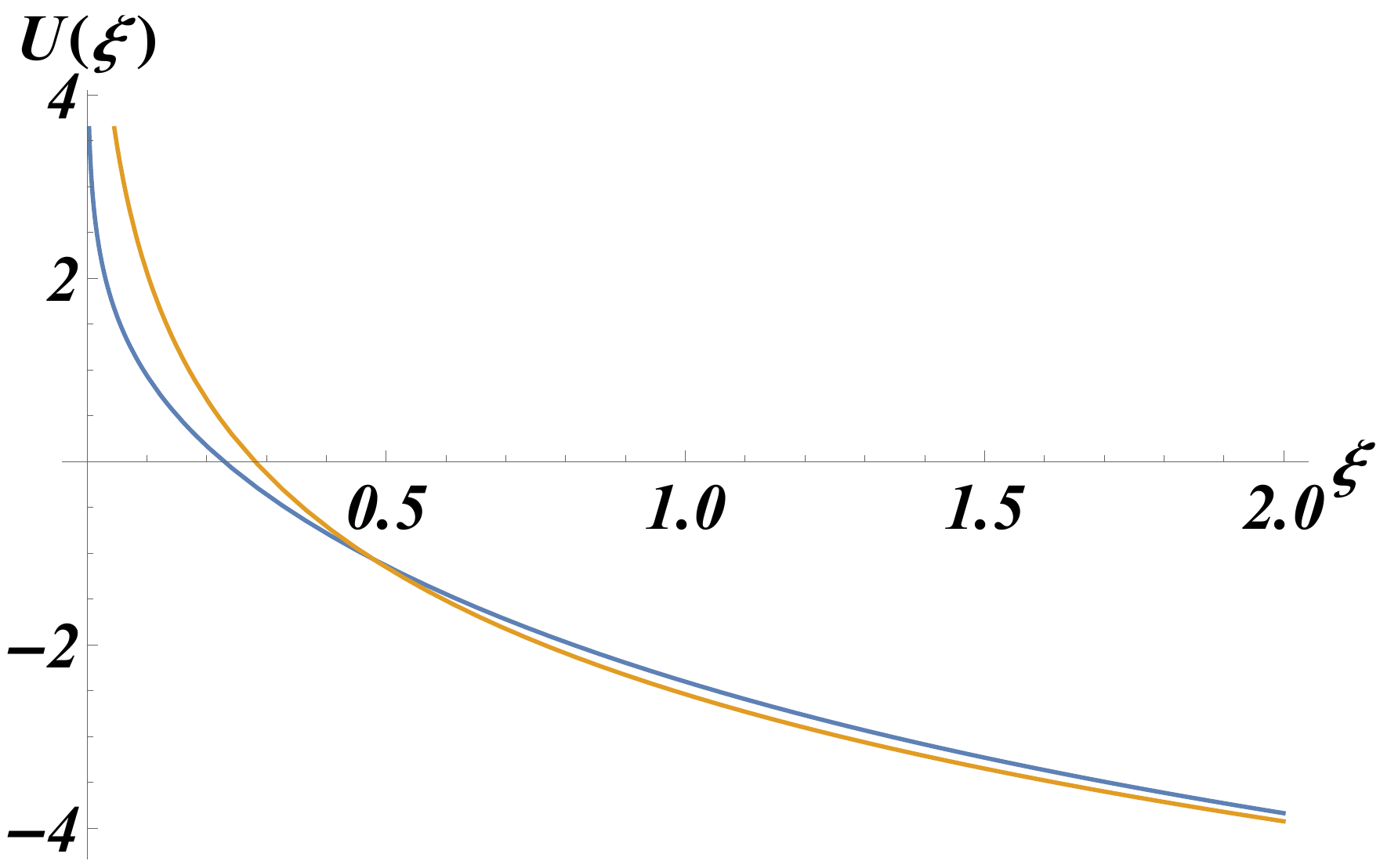}
    \caption{ Fourier transform (\ref{FourierTransform}) versus argument $\xi$ for the classical Enneper geometry with $n=2$ (blue color curve). It is also show a  comparison between numerical calculation of Eqs. (\ref{FourierTransform}),   and  analytical approximation (\ref{approxF}), corresponding to blue and orange curves, respectively.}
    \label{approxU}
\end{figure}

\paragraph{Classical Enneper case.} In order to isolated the singular part let us consider next approximation.  Firstly, let us focus on the classical Enneper surface $n=2$. One can argue, that  the most important contribution  to the integral is near $r=\epsilon$, where cubic term of the cosine argument may be neglected since $r^{3}\simeq \epsilon^{3}$, moreover,  for large $r$-value the contributions to the integral decay to zero as $r^{-1}$. Using this rationale, let us ignore the cubic term inside the argument of the cosine function, thus the integral reduces to the cosine integral
\begin{eqnarray}
\mathcal{U}(\xi)\simeq 2\int_{\epsilon}^{\infty}\frac{dr}{r}\cos\left(\xi r\right):=-2 {\rm Ci}(\epsilon \xi),
\end{eqnarray}
where ${\rm Ci}(x)$ is the cosine integral, that has the series expansion ${\rm Ci}(x)=\gamma +\log x+\sum_{n=1}^{\infty}(-1)^{n}x^{2n}/(2n (2n)!)$ \cite{gradshteyn2007}. Thus, one has
\begin{eqnarray}
\mathcal{U}(\xi)\simeq =-2\left(\gamma+\log\left(\xi\right)\right)+2\log\left(1/\epsilon\right)+\mathcal{O}(\epsilon^{2}),\label{approxF}
\end{eqnarray}
where $\gamma$ is the Euler-Mascheroni constant. Therefore the singular part is given by $2\log(1/\epsilon)$. Now, we performed numerical evaluation of (\ref{FourierTransform}), subtracting the singular part and comparing with previous result (\ref{approxF}) (see figure (\ref{approxU})). It is clear that the main error is close to the $k\simeq 0$, while for the rest of the values the error is around $1\%$. 

\begin{figure}[ht]
    \centering
       \includegraphics[scale=0.38]{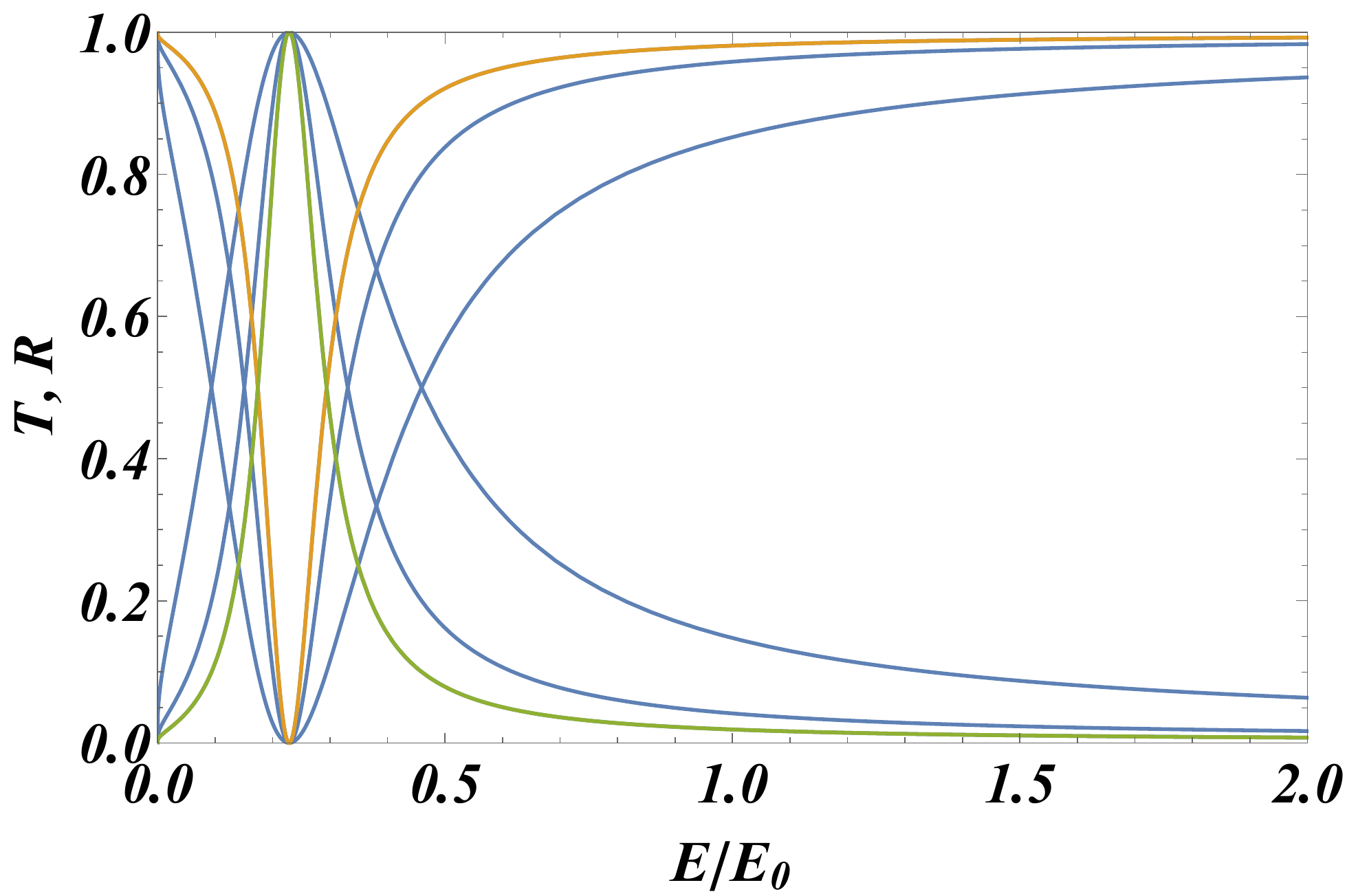}
    \caption{
Family of transmission and reflection coefficients (Eqs.(\ref{GenRes0}) and (\ref{GenRes}))  versus reduced energy $E_{*}=E/E_{0}$ using the numerical evaluation of the Fourier transform (\ref{FourierTransformBN}) corresponding to  the classical Enneper geometry with $n=2$. The set of curves was obtained for cases with $m=1, 2, 3$.  The orange and green curves are guides for the eyes to identify the $m=10$ case.   For all sets of transmission curves, we identify the same single Klein point $E_{K}\simeq 0.23 E_{0}$ where the transmission is $1$ and a clear trend of suppression after such point. 
        }
    \label{RefTransEnn}
\end{figure}

In figure (\ref{RefTransEnn}), it is shown the reflection and transmission coefficient for a propagation wave through the classical Enneper surface, where we have to compute numerically the Fourier transform $\mathcal{U}\left(\xi\right)$. 
The main feature of the transmission curves for the classical Enneper surface is the transmission value $\mathcal{T}(E_{K})=1$  at the energy $E_{K}$ showing the Klein paradox phenomena in this case \cite{DeCastro2001}. The Klein value $E_{K}$ corresponds to the value where the Fourier transform $\mathcal{U}(\xi)$ vanishes (see figure (\ref{approxU})). Using the approximation (\ref{approxF}) one can estimate the value of $E_{K}\simeq \frac{E_{0}}{2}e^{-\gamma}\approx 0.28 E_{0}$, with an error of $5\%$ respect to the numerical calculation. However,  in contrast to the previous cases, catenoid and $B_{1}$-Bour surfaces, in the present case, it is shown an strong suppression of the transmission for values above the Klein value $E_{K}$, giving rise to a total reflection effect and thus vanishing conductance $\mathcal{G}(E)\to 0$ for $E\gg E_{K}$. In addition, one can observe from figure (\ref{RefTransEnn}) that there are two values of interception and that both values moves towards  the Klein value as the $m$ value increases. This effect can be explain using the approximation (\ref{approxF}) by making the condition $m^{2}\left|\mathcal{U}(2E_{*}(m))\right|^{2}=1$, thus one obtain the values in terms of $m$ by the equation $E_{*}(m)\simeq E_{K}e^{\pm \frac{1}{2m}}$, showing the effect just mentioned.

\paragraph{B$n$-Bour cases for $n> 2$.}

In this case, one can follow the same line of argument as in the Enneper case. For instance, let us performed the change of variable $y=\frac{r^{n-1}}{n-1}$, thus Fourier transforms turns out as 
\begin{eqnarray}
\mathcal{U}\left(\xi\right)=\frac{2}{n-1}\int_{\epsilon}^{\infty}\frac{dy}{y} \cos\left(\xi \left(y+\alpha_{n}y^{\beta_{n}}\right)\right)\label{FourierTransformBN}, 
\end{eqnarray}
where $\alpha_{n}=(n-1)^{\beta_{n}}/(n+1)$ and $\beta_{n}=\frac{n+1}{n-1}$. Observe that $1<\beta_{n}\leq 3$ where equality corresponds to the classical Enneper case. Since $\beta_{n}>1$, one can attempt to argue that this term is not dominant near the singularity thus in this approximation one has $\mathcal{U}(\xi)\simeq-2{\rm Ci}(\epsilon\xi/(n-1))/(n-1)$, thus the difference between the present case and  the Enneper case  is a factor of $1/(n-1)$. However, in this case the error increase bigger than $10\%$. Thus, in this case we just compute the Fourier transform (\ref{FourierTransform}) numerically.

\begin{figure}[ht]
    \centering
       \includegraphics[scale=0.39]{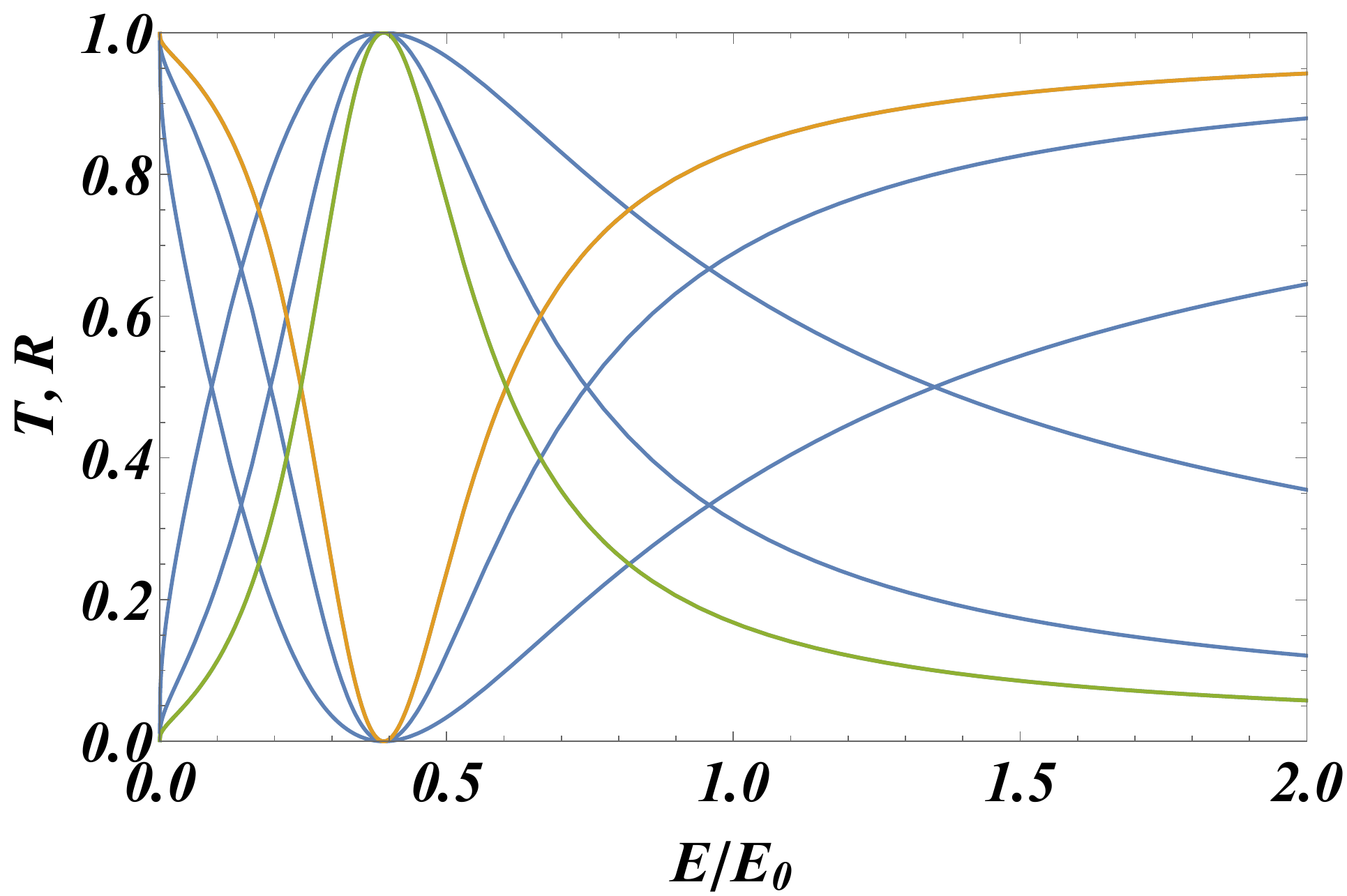}
        \includegraphics[scale=0.37]{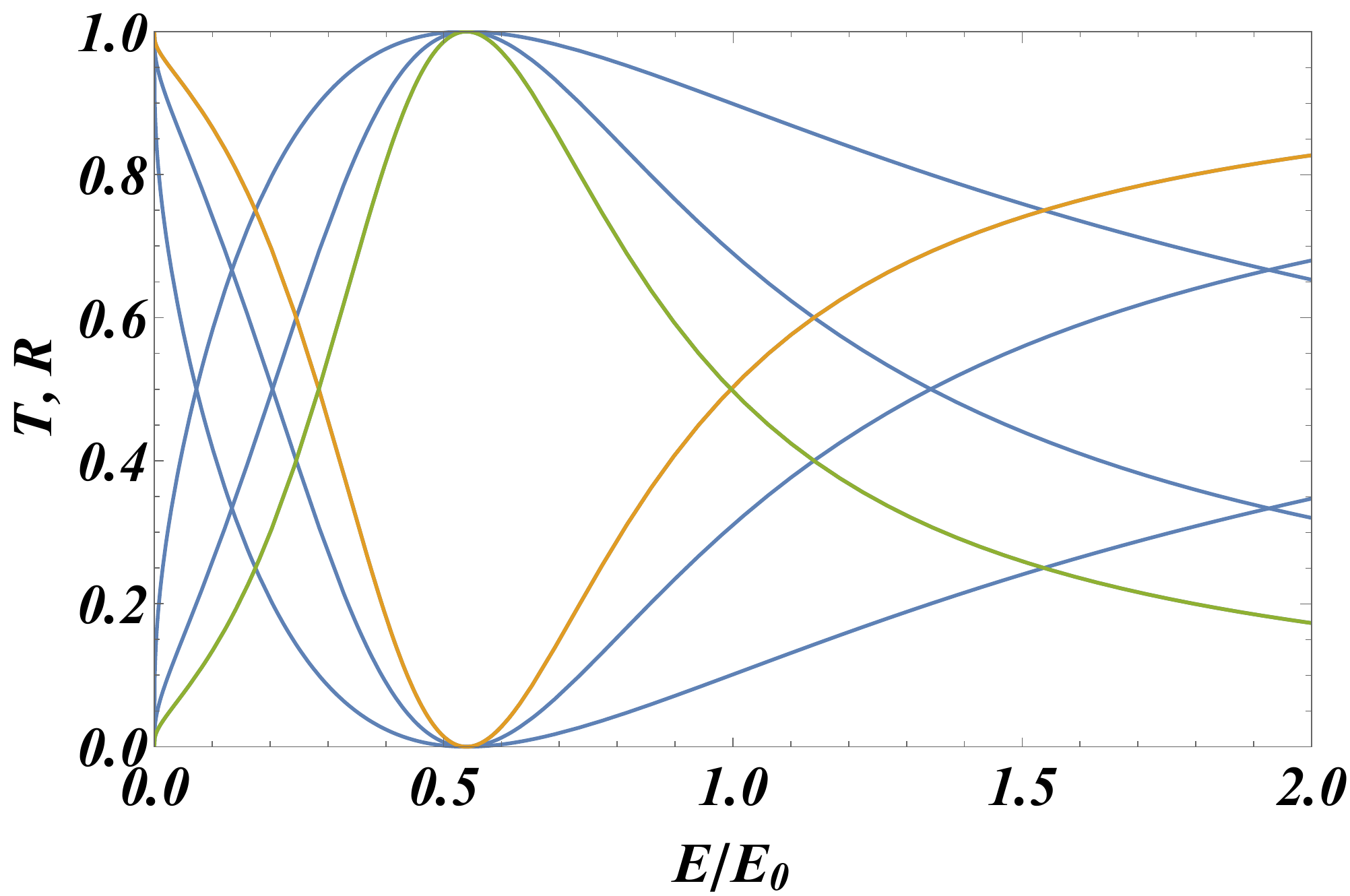}

    \caption{ Family of transmission and reflection coefficients (Eqs.(\ref{GenRes0}) and (\ref{GenRes}))  versus reduced energy $E_{*}=E/E_{0}$ using the numerical evaluation of the Fourier transform (\ref{FourierTransformBN}) corresponding to  the B3-Bour surface with $n=3$ (top) and  B4-Bour surface with $n=4$ (bottom). The set of curves was obtained for cases with $m=1, 2, 3$. The orange and green curves are guides for the eyes to identify the $m=3$ case. For all sets of transmission curves, we identify the same single Klein point $E_{K}\simeq 0.39 E_{0}$ (top) and $E_{K}\simeq 0.54 E_{0}$ (bottom), where the transmission is $1$, and a clear downward trend in transmission after that point. Moreover, the higher the value of $n$ is, the slower the downward trend is.
    }
    \label{RefTransB3Bour}
\end{figure}

In figures (\ref{RefTransB3Bour}), it is shown the reflection and transmission coefficient for a propagation wave through the $B_{3}$-Bour and $B_{4}$-Bour surfaces. Like in the classical Enneper case,  it can be appreciated that for each $B_{n}$-Bour surface there is a single Klein point $E_{K, n}$, where the transmittance is one. The Klein point moves to the right for greater values of $n$. After the Klein point $E_{K, n}$ the transmission decreases slowly as $n$ increases, however, it is also wholly suppressed for large values of energy.

\section{Concluding remarks}\label{sectVII}

In this article, we study the electronic degrees of freedom on a curved sheet of graphene-based on the Dirac equation. On this occasion we propose the hypothetical existence of a graphene sheet with the geometry of a Bour surface; examples of these surfaces are the catenoid, the helicoid and the classical Enneper surface, among others $B_{n}-$Bour surfaces that can be labelled using the $n$ parameter. Bour surfaces belong to the large family of minimal surfaces, those that minimize area or solutions of the Willmore shape equation. It is conspicuous that the geometry of the minimal surfaces was proposed to model specific carbon structures in \cite{Terrones}. Although there is still not an artificial or natural realization of these carbon allotropes in either laboratory or nature, there are good expectations of their existence from numerical and experimental investigations \cite{Terrones2, Terrones3, Braun2018, Tanabe2020},.

Now, for each $n$, the space-time $\mathbb{M}$ is built with the global structure of $\mathbb{M}=\mathbb{R}\times{B_{n}}$
over which we define the Dirac field. In particular, by means of an elementary change of parameters, it is possible to rewrite the metric of $\mathbb{M}$ as $ds^{2}=-v^{2}_{F}dt^{2}+dx^{ 2}+\left(1/V^{2}\left(x\right)\right)d\theta^{2}$, and the Dirac equation as \begin{eqnarray}
i\hbar\partial_{t}\Phi=v_{F}\sigma_{1}\hat{p}_{x}\Phi+v_{F}V(x)\sigma_{2}\hat{\ell }_{\theta}\Phi,\label{DiracEqBourCon}
\end{eqnarray}
where $v_{F}$ is the Fermi velocity, being $\sigma_{1}$ and $\sigma_{2}$ the Pauli matrices. In this equation, $V(x)$ has been  interpreted as an effective scattering potential coupled to a pseudo-spin orbit term of the form $\sigma_{2}\ell_{\theta}$, where $\sigma_ {2}$ the direction of the pseudo-spin and $\hat{\ell}_{\theta}$ the angular momentum in two dimensions. For each Bour surface,  it was found that $V(x)$ decays to zero as $x\to\infty$, while $V(x)$ works as a potential barrier near $x=0$. In fact, it can be shown that for $n\geq 2$, $V(x)$ approach to a repulsive Coulombic type potential.

The asymptotic behaviour of the states in $x\to\infty$ is determined with (\ref{DiracEqBourCon}), which effectively corresponds to solutions of a Dirac equation in a space-time $1+1$. These states in terms of $x$ are characterised as plane waves with a pseudo-spin up $\uparrow$ (or down $\downarrow$) depending on the positive or negative value of the energy. Furthermore, through the Lippmann-Schwinger formalism, we studied the out-scattering states,  giving rise to an out-scattering state divided into a transmitted and a reflected wave. This is done through the Born and the high-order  Born approximation, which can be summed up. In particular, it is observed that the reflected wave transmutes the pseudo-spin direction, which we coined the spin-orbit interaction.

In addition, through the  N$\ddot{\rm o}$ether current $J^{\mu}$,  the probability density, $J^{0}$, is determined, which allows us to argue that it is more probable to find Dirac fermions near the scattering point, in fact, within this approximation we found that the probability density is proportional to $V(x)$. Now, using the spatial components of $J^{\mu}$, the incident and scattered currents are determined to find expressions for the reflectance $\mathcal{R}(E)$, and the transmittance $\mathcal{ T}(E)$, respectively. It is found that for the Bour surfaces $B_{0}$, catenoid (or helicoid), and $B_{1}$-Bour, there is usual behaviour for the transmittance and reflectance, giving rise to the effect of total transmittance for large values of energy. Although the potential barrier in the cases $B_{0}$ and $B_{1}$ evokes the usual situation where Klein's paradox arises, the difference lies in the coupling with $\sigma_{2}$ that appears in (\ref{DiracEqBourCon}), which we coin the absence of Klein's paradox \cite{DeCastro2001}. However, for Bour surfaces $B_{n}$ with $n\geq 2$, including the classical Enneper surface, we show that there is an energy point $E_{K}$ for which the transmittance is equal to $\mathcal{T}(E_{K})=1$, giving rise to a manifestation of Klein's paradox, while for large values of energy $E\gg E_{K}$ the transmittance decays to zero suppressing the conductance completely.

The present work can be extended as follows. For the $n\geq 2$ one can approximate the geometry-induced potential $V(x)$ as a Coulombic potential near the scattering region, where one can attempt to figure out an analytical solution for the states and the electronic spectrum. Following a different direction, through the Weierstrass-Enneper representation, we can propose the study of electronic degrees of freedom on other minimal surfaces such as simply periodic minimal surfaces, k-noids or Schwartzites that are much more involved. In particular, for these surfaces, it is found that the conformal factor $\Lambda(r, \theta)$ depends intricately on $r$ and $\theta$, so it is not possible to perform a separation as in the case of Bour surfaces \cite{Ulrich}. However, we can implement traditional methods like the finite element to solve the Dirac equation (\ref{DiracPolar}) to study other electronic properties like the density of states, Kubo conductivity, and so on.

\section*{Acknowledgement}

V. A. G-D. acknowledges the financial support provided by
Conacyt (No. CVU 736886). P. C.-V.  would like to thank Idrish Huet-Hern\'andez and Romeo de Coss for many valuable discussions.

\appendix
\section{Green function calculation}\label{AppA}

Calculation of the one-dimensional Green function. Let us start from the equation 
$\left(E-v_{F}\sigma_{1}\hat{p}_{x}\right)\mathbb{G}\left(x, x^{\prime}, E\right)=\delta\left(x-x^{\prime}\right)$, where the momentum operator $\hat{p}_{x}=-i\hbar \partial_{x}$ and let us recall that the dispersion relation is given by $E=\pm \hbar v_{F}\left|k\right|$. Now, let us define the function $g_{0}\left(x, x^{\prime}, E\right)$ such that $\mathbb{G}\left(x, x^{\prime}, E\right)=\left(E+v_{F}\sigma_{1}\hat{p}_{x}\right)g_{0}\left(x, x^{\prime}, E\right)$, thus it is not difficult to show that $g_{0}\left(x, x^{\prime}, E\right)$ satisfies the Green-Helmholtz equation 
$\left(\hbar v_{F}\right)^{2}\left(-\partial_{x}^{2}+k^2\right)g_{0}\left(x, x^{\prime}, E\right)=\delta(x-x^{\prime})$.
Now, the solution of this equation is known to be \cite{morse_feshbach_1981}
\begin{eqnarray}
g_{0}\left(x, x^{\prime}, E\right)=\frac{1}{2i(\hbar v_{F})^{2}\left|k\right|}e^{i\left|k\right|\left|x-x^{\prime}\right|}.
\end{eqnarray}

 \begin{widetext}
 
\section{Calculation of the higher-order Born approximation}\label{AppB}
  \subsection{Calculation of $\tau_{n+1}(p)$ terms}

 We start with the expression (\ref{termkm1})
\begin{eqnarray}
\boldsymbol{\tau}_{n+1}\left(p\right)&=&(\hbar v_{F})^{n+1}(2\pi)^{n}\sum_{{\bf q}^{(1)}, \cdots, {\bf q}^{(n)}}\sigma_{2}m\delta_{mm^{(1)}}\tilde{U}\left(p-q^{(1)}\right)\left(\prod_{\ell=1}^{n-1}\mathbb{G}({ q}^{(\ell)})\sigma_{2}m^{(\ell)}\delta_{m^{(\ell)}m^{(\ell+1)}}\tilde{U}\left(q^{(\ell)}-q^{(\ell+1)}\right)\right)\nonumber\\
&\times&  \sigma_{2}v_{\mu\cdot\sigma}f_{m^{(n)}}m^{(n)}\tilde{U}\left(q^{(n)}-\sigma\left|k\right|\right),\label{termkm1s}
\end{eqnarray}
where we have substituted the expression for type {\it a} (\ref{typea}) and type {\it b} (\ref{typeb}). Taking advantage of the Kronecker deltas $\delta_{m^{(\ell)}m^{(\ell+1)}}$ we are able to simplify the last expression as follows (note that each $2\pi$ cancels out with each $2\pi$ that appears in $\frac{1}{2\pi}\sum_{m}$)

\begin{eqnarray}
\boldsymbol{\tau}_{n+1}\left(p\right)&=&(\hbar v_{F}m)^{n+1}f_{m}
\int\left(\prod_{\ell=1}^{n}\frac{dq^{(\ell)}}{2\pi}\right)\sigma_{2}\tilde{U}\left(p-q^{(1)}\right)\left(\prod_{\ell=1}^{n-1}\mathbb{G}({ q}^{(\ell)})\sigma_{2}\tilde{U}\left(q^{(\ell)}-q^{(\ell+1)}\right)\right)\mathbb{G}({ q}^{(n)})\nonumber\\
&\times&
\sigma_{2}v_{\mu\cdot\sigma}\tilde{U}\left(q^{(n)}-\sigma\left|k\right|\right).
\end{eqnarray}
Now, we organize the integrals in the following nested structure
\begin{eqnarray}
\boldsymbol{\tau}_{n+1}\left(p\right)&=&(\hbar v_{F}m)^{n+1}f_{m}
\int \frac{dq^{(1)}}{2\pi}\sigma_{2}\tilde{U}\left(p-q^{(1)}\right)\mathbb{G}\left({ q}^{(1)}\right)\mathcal{U}^{(1)}\left(q^{(1)}\right)\sigma_{2}v_{\mu\cdot\sigma},
\nonumber\\
\label{ddtermkm1}
\end{eqnarray}
where $\mathcal{U}^{(1)}(q^{(1)})$ is written in terms of $\mathcal{U}^{(2)}(q^{(2)})$, and so on. In general, one has the following definition
\begin{eqnarray}
\mathcal{U}^{(\ell)}\left(q^{(\ell)}\right)=\int \frac{dq^{(\ell+1)}}{2\pi}\sigma_{2}\tilde{U}\left(q^{(\ell)}-q^{(\ell+1)}\right)\mathbb{G}\left({ q}^{(\ell+1)}\right)\mathcal{U}^{(\ell+1)}\left(q^{(\ell+1)}\right), \nonumber
\end{eqnarray}
where $\ell=1, \cdots, n-1$ and $\mathcal{U}^{(n)}(q^{(n)})=\tilde{U}\left(q^{(n)}-\sigma\left|k\right|\right)$.

Next, let us proceed to calculate $\mathcal{U}^{(n-1)}(q^{(n-1)})$. The integral involved in this quantity can be performed using complex integration replacing $q^{(n)}$ by the complex variable $z$
\begin{eqnarray}
\mathcal{U}^{(n-1)}(q^{(n-1)})=i\int_{\Gamma_{1}} \frac{dz}{2\pi i}\sigma_{2}\tilde{U}\left(q^{(n-1)}-z\right)\mathbb{G}\left(z\right)\tilde{U}\left(z+\left|k\right|\right),\label{intz}
\end{eqnarray}
 where we have  put $\sigma=-1$ since we have an  initial left wave.  The contour complex integration $\Gamma_{1}$ is chosen as it is shown in the left side of  figure (\ref{fig3}) since we exclude the points where the argument of the Fourier transform is zero. Note that $\tilde{U}(0)=\int_{-\infty}^{\infty}dx U(x)$ is strictly divergent since $U(x)$ is a long-range potential which for all Bour surfaces decays as $1/x$. 
\begin{figure}[ht]
    \centering
    \includegraphics[scale=0.5]{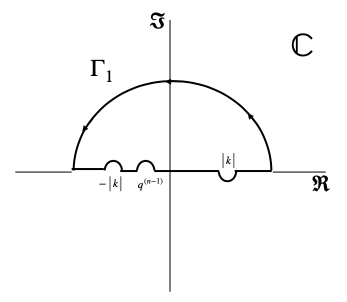}~~~~~~~~~~   \includegraphics[scale=0.5]{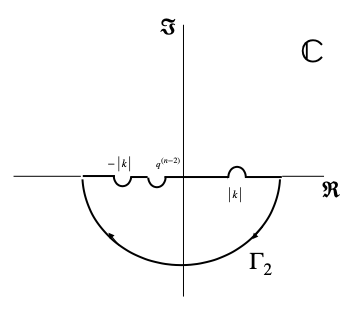}
    \caption{Contour curves $\Gamma{1}$ and $\Gamma{2}$ for the complex integrals in the $z$  plane  (Eq. (\ref{intz})) and in the $q^{n-1}$  plane (Eq. (\ref{intqnm1})), respectively.}
    \label{fig3}
\end{figure}
The Green function $\mathbb{G}\left(p\right)$ in momentum space can be written as
\begin{eqnarray}
\mathbb{G}(p)=\frac{\mu}{\hbar v_{F}}\frac{ \left|k\right|+\mu\sigma_{1}p}{\left[p-\left(\left|k\right|+i\epsilon\right)\right]\left[p+\left(\left|k\right|+i\epsilon\right)\right]}\nonumber
\end{eqnarray}
where $\mu=\pm$ represents the positive and negative energy states, and one can identify two poles at $\left|k\right|+i\epsilon$ and $-\left|k\right|-i\epsilon$. Using the Cauchy integral theorem it is not difficult to show that 
\begin{eqnarray}
\mathcal{U}^{(n-1)}(q^{(n-1)})=\tilde{U}\left(q^{(n-1)}-\left|k\right|\right)\frac{i\mu}{\hbar v_{F}}\tilde{U}\left(2\left|k\right|\right)\sigma_{2}\mathbb{P}_{\mu}\nonumber
\end{eqnarray}
where $\mathbb{P}_{\mu}=\frac{1}{2}\left(1+\mu \sigma_{1}\right)$ is a projector.  Now, in order to be transparent in the calculation, let us insert this result into the integration by the $q^{(k-1)}$ variable, turning it out as
{\small \begin{eqnarray}
\mathcal{U}^{(n-2)}(q^{(n-2)})&=&\int \frac{dq^{(n-1)}}{2\pi}\sigma_{2}\tilde{U}\left(q^{(n-2)}-q^{(n-1)}\right)\mathbb{G}\left({ q}^{(n-1)}\right)\tilde{U}\left(q^{(n-1)}-\left|k\right|\right)\frac{i\mu}{\hbar v_{F}}\tilde{U}\left(2\left|k\right|\right)\sigma_{2}\mathbb{P}_{\mu}\nonumber
\end{eqnarray}}
Although,  the integration is similar to the previous one, it is convenient to perform the change of variable $q^{(n-1)}\to -q^{(n-1)}$, thus the integration results in 
{\small\begin{eqnarray}
\mathcal{U}^{(n-2)}(q^{(n-2)})&=&\int \frac{dq^{(n-1)}}{2\pi}\sigma_{2}\tilde{U}\left(q^{(n-2)}+q^{(n-1)}\right)\mathbb{G}\left(-{ q}^{(n-1)}\right)\tilde{U}^{*}\left(q^{(n-1)}+\left|k\right|\right)\frac{i\mu}{\hbar v_{F}}\tilde{U}\left(2\left|k\right|\right)\sigma_{2}\mathbb{P}_{\mu},\label{intqnm1}
\end{eqnarray}}
where $\tilde{U}^{*}(k)=\tilde{U}(-k)$ is the complex conjugate. 
Now, we proceed to calculate this integral using again complex integration as the previous integration. The result is the same except that $\mathbb{P}_{\mu}$ changes by $\mathbb{P}_{-\mu}$, that is, 
\begin{eqnarray}
\mathcal{U}^{(n-2)}(q^{(n-2)})=\tilde{U}\left(q^{(n-2)}+\left|k\right|\right)\left(\frac{i\mu}{\hbar v_{F}}\right)^2\left|\tilde{U}\left(2\left|k\right|\right)\right|^{2}\sigma_{2}\mathbb{P}_{-\mu}\sigma_{2}\mathbb{P}_{\mu}, \nonumber
\end{eqnarray}
where $\left|\cdot\right|$ is the complex norm.  Now, by an iterative process one can conclude that 
\begin{eqnarray}
\mathcal{U}^{(1)}(q^{(1)})=\tilde{U}\left(q^{(1)}+(-1)^{n-1}\left|k\right|\right)\left(\frac{i\mu}{\hbar v_{F}}\right)^{n-1}
\mathbb{B}^{(n)}_{\mu},
\nonumber
\end{eqnarray}
where 
\begin{eqnarray}
\mathbb{B}^{(n)}_{\mu}=\left\{\begin{array}{cc}
\prod_{j=1}^{(n-1)/2}\left[\left|\tilde{U}\left(2\left|k\right|\right)\right|^{2}\sigma_{2}\mathbb{P}_{-\mu}\sigma_{2}\mathbb{P}_{\mu}\right], & {\rm for}~~n~~{\rm odd},\\
&\\
\tilde{U}\left(2\left|k\right|\right)\sigma_{2}\mathbb{P}_{\mu}\prod_{j=1}^{(n-2)/2}\left[\left|\tilde{U}\left(2\left|k\right|\right)\right|^{2}\sigma_{2}\mathbb{P}_{-\mu}\sigma_{2}\mathbb{P}_{\mu}\right], & {\rm for}~~n~~{\rm even}.
\end{array}\right.
\label{B}
\end{eqnarray}
Note that (\ref{B}) can be  simplified further as a consequence of the following algebra $\sigma_{2}\mathbb{P}_{\mu}=\mathbb{P}_{-\mu}\sigma_{2}$, thus $\sigma_{2}\mathbb{P}_{-\mu}\sigma_{2}\mathbb{P}_{\mu}=\mathbb{P}_{\mu}\sigma_{2}^{2}\mathbb{P}_{\mu}=\mathbb{P}_{\mu}^{2}=\mathbb{P}_{\mu}$.  Using this algebra and the property $\mathbb{P}_{\mu}^2=\mathbb{P}_{\mu}$, one obtain 
\begin{eqnarray}
\mathbb{B}^{(n)}_{\mu}=\left\{\begin{array}{cc}
\left|\tilde{U}\left(2\left|k\right|\right)\right|^{n-1}\mathbb{P}_{\mu}, & {\rm for}~~n~~{\rm odd},\\
&\\
\tilde{U}\left(2\left|k\right|\right)\left|\tilde{U}\left(2\left|k\right|\right)\right|^{n-2}\sigma_{2}\mathbb{P}_{\mu}, & {\rm for}~~n~~{\rm even}.
\end{array}\right.
\nonumber
\end{eqnarray}

Now, we substitute $\mathcal{U}^{(1)}(q^{(1)})$ inside the expression for $\tau_{n+1}(p)$ in equation (\ref{ddtermkm1}), and we proceed to perform the calculation for $n$ odd and even cases using the same estrategy used to calculate the integrals on the variables $q^{(k)}$ and $q^{(k-1)}$. Additionally, we use the property $\mathbb{P}_{\mu}v_{\mu}=v_{\mu}$ and $\sigma_{2}v_{-\mu}=i\mu v_{\mu}$.  Thus the result of the $n+1-{\rm th}$ term corresponds to the  expressions in (\ref{tauoddn}) and (\ref{tauevenn}). 

\end{widetext}

 \subsection{Calculation of $C_{n+1}(x)$ terms}

The starting point to calculate the terms $C_{n+1}(x)$ corresponds to the Fourier integral
\begin{eqnarray}
C_{n+1}(x)=\int \frac{dp}{2\pi}e^{ipx}\mathbb{G}(p)\tau_{n+1}\left(p\right), \nonumber
\end{eqnarray}
in the cases even $n$  and odd $n$. For the odd $n$, we use the equation (\ref{tauoddn}),  make the change of variable $p\to-p$, and perform the complex integration using the contour $\Gamma_{2}$, while that for the even $n$, we use the equation (\ref{tauevenn}), and the complex integration is performed using the contour $\Gamma_{1}$. In this manner, we obtain the desired expressions (\ref{Coddj})
  and (\ref{Cevenj}), respectively.


\section{Natural coordinates Catenoid}\label{AppC}

\subsubsection{Cartesian coordinates.} The metric of the space-time $\mathbb{M}$ written through the square of the line element considered here is 
\begin{eqnarray}
ds^2=-v_{F}^2 dt^2+\Lambda^2\left(\omega\right)\left|d\omega\right|^2,\label{SpaceTime}
\end{eqnarray}
where $\Lambda^2\left(\omega\right)$ is the conformal factor introduced above for the minimal surfaces, and $\left|d\omega\right|^2=du^2+dv^2$. The local indices in this case can be split as $\alpha=t, u,v$. From the metric (\ref{SpaceTime}) one can easily read $\hat{\theta}^{0}=v_{F}dt$, $\hat{\theta}^{1}=\Lambda du$ and $\hat{\theta}^{2}=\Lambda dv$, from where one can extract the components of the vielbeins $e^{A}_{\mu}$. Now, from the Maurer-Cartan equation (\ref{Maurer1}) and the torsionless condition one can obtain $d\hat{\theta}^{0}=0$, and 
\begin{eqnarray}
d\hat{\theta}^{1}+\frac{\Lambda_{v}}{\Lambda^{2}}\hat{\theta}^{1}\wedge \hat{\theta}^{2}&=&0\label{MCCat1}\\
d\hat{\theta}^{2}+\frac{\Lambda_{u}}{\Lambda^{2}}\hat{\theta}^{2}\wedge \hat{\theta}^{1}&=&0\label{MCCat2}\\
\end{eqnarray}
Now, from the equation (\ref{MCCat1}) one can deduce $\omega\indices{^{1}_{0}}=0$ and $\omega\indices{^{1}_{2}}=\frac{\Lambda_{v}}{\Lambda^2}\hat{\theta}^{1}+X\hat{\theta}^{2}$ for some local function $X$, whereas from (\ref{MCCat2}) one can deduce that  $\omega\indices{^{2}_{0}}=0$ and $\omega\indices{^{2}_{1}}=\frac{\Lambda_{u}}{\Lambda^2}\hat{\theta}^{2}+\tilde{X}\hat{\theta}^{1}$. Now, we use the metric condition, $\omega^{AB}=-\omega^{BA}$, thus one can determine $X$ and $\tilde{X}$, turning that the only non-zero components of the connection one-form are 
\begin{eqnarray}
\omega^{12}=-\omega^{21}=\frac{\Lambda_{v}}{\Lambda^2}\hat{\theta}^{1}-\frac{\Lambda_{u}}{\Lambda^2}\hat{\theta}^{2}
\end{eqnarray}
 
These components expressed in local coordinates are given by
$\omega\indices{_{u}^{12}}=-\omega\indices{_{u}^{21}}=\partial_{v}\log\Lambda\left(\omega\right)$ and $\omega\indices{_{v}^{12}}=-\omega\indices{_{v}^{21}}=-\partial_{u}\log\Lambda\left(\omega\right)$.  Consequently, the spin connection $\Omega_{\alpha}$ is given simply as $\Omega_{t}=0$, $\Omega_{u}=\frac{i}{2}\partial_{v}\log\Lambda\left(\omega\right)\sigma_{3}$, and $\Omega_{v}=-\frac{i}{2}\partial_{u}\log\Lambda\left(\omega\right)\sigma_{3}$.

Now, we use all these information in order to write down an explicit expression for the Dirac equation in these space-times. Denoting the $2+1$ Dirac spinor by $\Psi$ and making the transformation $\Psi=\Lambda^{-\frac{1}{2}}\Phi$, we are able to show that the Dirac equation is given by 
\begin{eqnarray}
i\hbar\partial_{t}\Phi=-i\frac{\hbar v_{F}}{\Lambda}\left(\sigma_{1}\partial_{u}\Phi+\sigma_{2}\partial_{v}\Phi\right).
\label{Dirac2}
\end{eqnarray}
Clearly, in the simplest case when $\Lambda=1$ the above equations correspond to the Dirac equation in Minkowski's space-time. The Dirac equation in these coordinates $u,v$ is 
particularly useful in the case when the conformal factor $\Lambda$ depends on one of the coordinates. Noticeably, the equation (\ref{Dirac2}) is valid for any conformally flat space metric \cite{Cvetic}.

\subsection{Dirac equation  on the catenoid using natural coordinates}
In this section, we write down the Dirac equation in the most natural coordinates of the catenoid before to consider the Weierstrass-Enneper representation (\ref{WErep}). Indeed, let us consider the parametrization of the catenoid obtained from the $2\pi-$rotation of the catenary, that is,
\begin{eqnarray}
{\bf X}(z, \varphi)=\left({R}(z)\cos\varphi, -{R}(z)\sin\varphi, z\right),
\end{eqnarray}
where ${\bf R}(z)=R_{0}\cosh(z/R_{0})$, with $R_{0}$ the radius of the neck of the catenoid, where $z\in \left(-\infty, \infty\right)$, and $\varphi\in\left[0,2\pi\right)$. The metric square line in this case is given by 
\begin{eqnarray}
ds^2=R_{0}^2\cosh^2\left(\frac{z}{R_{0}}\right)\left(\frac{1}{R^2_{0}}dz^2+d\varphi^2\right).
\end{eqnarray}
Clearly, one can identify the coordinates $u\to \zeta=z/R_{0}$ and $v\to \varphi$, and the conformal factor $\Lambda(\omega)\to \lambda(\zeta)=R_{0}\cosh(z/R_{0})$. The expression (\ref{Dirac2}) turn out particularly useful  since the conformal factor just depends on one of the two coordinates.  Further, it is convenient to defined the following change of variables $x=R_{0}\sinh\zeta$, where $x\in\left(-\infty,\infty\right)$, thus it is not difficult to show that $\lambda^{-1}(\zeta)\partial_{\zeta}=\partial_{x}$.
Indeed, the Dirac equation reduces to the  equation found above (\ref{DiracEqBour}), $i\hbar\partial_{t}\Phi=v_{F}\sigma_{1}\hat{p}_{x}\Phi+v_{F}\sigma_{2}V(x)\hat{\ell}_{\varphi}\Phi$, where $\hat{p}_{x}=-i\hbar\partial_{x}$ and $\hat{\ell}_{\varphi}=-i\hbar\partial_{\varphi}$ are linear  and angular momentum operators; and the expected effective potential found above $V(x)=1/\sqrt{x^2+R^{2}_{0}}$. The connection between the natural coordinates and the polar coordinates can be accomplished by using the change of  variable $r=e^{\zeta}$ and identified $\varphi\to\theta$.

\begin{widetext}
\section{Explicit parametrizations}
In this section, we show explicit parametrizations for the Catenoid, Helicoid, $B_{1}-$Bour and the classical Enneper surfaces in polar coordinates $(r,\varphi)$, following \cite{Whittemore}. For the catenoid
{\small\begin{eqnarray}
{\bf X}_{c}(r, \varphi)=\left(\frac{\alpha}{2}\left(\frac{1}{r}+r \right)\cos{\varphi}, \frac{\alpha}{2}\left(\frac{1}{r}+r \right)\sin{\varphi}, -\alpha\log{r}\right);\nonumber\\
\label{ParamCat}
\end{eqnarray}}
for the helicoid
{\small\begin{eqnarray}
{\bf X}_{h}(r, \varphi)=\left(\frac{\beta}{2}\left(r - \frac{1}{r} \right)\sin{\varphi}, \frac{\beta}{2}\left(r-\frac{1}{r} \right)\cos{\varphi}; \beta \varphi\right).\nonumber\\
\label{ParamHelicoid}
\end{eqnarray}}
for the classical Enneper surface
{\small\begin{eqnarray}
{\bf X}_{e}(r, \varphi)=\left(r \cos{\varphi}- \frac{r^3}{3}\cos{3\varphi}, -r \sin{\varphi} - \frac{r^3}{3}\sin{3\varphi}, r^2 \cos{2\varphi}\right)\nonumber\\
\label{ParamEnneper};
\end{eqnarray}}
and for the $B_{1}-$Bour surface
{\small\begin{eqnarray}
{\bf X}_{b}(r, \varphi)=\left(\log{r} - \frac{r^2}{2}\cos{2\varphi}, -\varphi - \frac{r^2}{2}\sin{2\varphi}, 2r \cos{\varphi}\right)\nonumber\\
\label{ParamB1},
\end{eqnarray}}
Using these parametrizations, ${\bf X}_{c}, {\bf X}_{h}, {\bf X}_{b}$ and ${\bf X}_{e}$, it is drawn the surfaces inside the figures (\ref{fig1}), (\ref{fig22}) and (\ref{fig223})) with help of {\it Mathematica}. 
\end{widetext}
\nocite{*}

 \bibliography{Referencias.bib}

\end{document}